\documentclass[
12pt,
superscriptaddress,
amsmath,amssymb
]{revtex4}

\usepackage{graphicx}
\usepackage{subfigure}
\usepackage{dcolumn}
\usepackage{bm}
\usepackage{multirow}
\usepackage{mathrsfs}
\usepackage{xcolor}

\newcommand{\bea}{\begin{eqnarray}}
\newcommand{\eea}{\end{eqnarray}}
\newcommand{\be}{\begin{equation}}
\newcommand{\ee}{\end{equation}}

\newcommand{\ket}[1]{\vert{#1}\rangle}
\newcommand{\veps}{\varepsilon}
\newcommand{\Vc}{\mathcal{V}}
\newcommand{\Tc}{\mathcal{T}}

\begin{document}

\title{Calculating eigenvalues and eigenvectors of parameter-dependent hamiltonians using an adaptative wave operator method.}
\author{Arnaud Leclerc}
\email{Arnaud.Leclerc@univ-lorraine.fr}
\affiliation{Universit\'e de Lorraine, CNRS, Laboratoire de Physique et Chimie Th\'eorique, UMR7019, F-57000 Metz, France}

\author{Georges Jolicard}
\affiliation{Institut UTINAM UMR CNRS 6213, Observatoire de Besan\c{c}on, 25010 Besan\c{c}on Cedex, France}

\begin{abstract}

We propose a wave operator method to calculate eigenvalues and eigenvectors of large parameter-dependent matrices, using an adaptative active subspace. 
We consider a hamiltonian which depends on external adjustable or adiabatic parameters, using adaptative projectors which follow the successive eigenspaces when the adjustable parameters are modified. 
The method can also handle non-hermitian hamiltonians. 
An iterative algorithm is derived and tested through comparisons with a standard wave operator algorithm using a fixed active space and with a standard block-Davidson method. 
The proposed approach is competitive, it converges within a few dozen iterations at constant memory cost. 
We first illustrate the abilities of the method on a 4-D coupled oscillator model hamiltonian. 
A more realistic application to molecular photodissociation under intense laser fields with varying intensity or frequency is also presented. Maps of photodissociation resonances of H${}_2^+$ in the vicinity of exceptional points are calculated as an illustrative example. 

\end{abstract}
\maketitle

\section{Introduction \label{introduction}}

There are many problems in molecular physics which can be described by parametric hamiltonians. 
A parameter-dependent hamiltonian generally arises in the context of adiabatic separations when a ``slow" coordinate is considered as a parameter to solve the eigenvalue problem associated with another ``rapid" coordinate. 
The most emblematic example is the Born-Oppenheimer approximation
 in molecular calculations, 
 where the mass ratio between nuclei and electrons allows for an adiabatic separation of their respective coordinates. 
The slow nuclear coordinates become parameters in the Schr\"odinger equation for the fast electrons\cite{bornoppenheimer,liehr1957,messiahchXVIII}. 
The electronic Hamiltonian is thus parameter-dependent, the parameters being here the coordinates of the nuclei, and one has to diagonalize it for many different values of the parameters. 
The Born-Oppenheimer approximation can also be used to separate fast and slow nuclei motions in weakly bound molecular complexes to calculate bound states   \cite{holmgren1977,frey1985,zeng2011,leforestier2012}
or to perform scattering calculations \cite{scribano2012}. 
In those situations the hamiltonian for the fast nuclei is also parameter-dependent, the parameters being the coordinates of the slow relative motion of interacting molecules. 

An even more prominent motivation for studying parameter-dependent hamiltonians comes from the quantum control context where molecules are illuminated by strong laser fields\cite{bookshapiro2003,bookshore2011}. 
We can think about the Floquet hamiltonian of a molecule submitted to laser pulses whose dynamics is described by quasienergy states \cite{shirley,reviewguerin2003}. 
In this case, the adiabatic parameters arising in the hamiltonian correspond to the intensity and/or frequency of the laser field used to control the molecule. 
In this paper, 
we are precisely motivated by this kind of problems, among which we shall focus on H$_2^+$ photodissociation as a representative example (although the proposed method could in principle fit all the situations suggested above). 
In the H$_2^+$ case, an extensive knowledge of the quasienergy landscape can be very useful to design efficient adiabatic control strategies taking advantage of some peculiarities in the spectrum, 
such as exceptional points (\textit{i.e.} non-hermitian degeneracies)\cite{moiseyevNHQM,lefebvre2009,lefebvre2012}
or zero-width resonances (\textit{i.e.} infinitely long-lived resonances)\cite{atabek2006,atabek2013,leclerc2016,leclerc2017}. 
%

In all the physical situations described above, 
one has to diagonalize not only \emph{one} hamiltonian, but a \emph{series} of hamiltonians corresponding to different  values of the external parameters. 
Keeping in mind the laser control example, we propose to design a specific algorithm to efficiently calculate eigenvalues and eigenvectors of parameter-dependent hamiltonians, in which we take into account the fact that successive hamiltonians are close from one another when the external parameter(s) is(are) modified. 
%
We shall consider the following generic, parametric eigenvalue problem, 
\be  
H(\veps) \Vc(\veps) = \Vc(\veps) E(\veps) 
\ee 
where $H$ is the $N\times N$ matrix representation of some hamiltonian operator depending on one or several continuous parameters denoted collectively by $\veps$. $\Vc$ is the $N\times N$ matrix of 
 eigenvectors in columns and $E$ the diagonal matrix of eigenvalues $E_i, i=1,\dots,N$. 
Actually $H$ can be any parametric matrix, not only the matrix representation of a hamiltonian operator. It can be hermitian or not. If not, we will restric ourselves to complex symmetric matrices in the applications and focus on the right eigenvectors.

In what follows, we assume that the parameter follows a monotonic variation and successively takes discretized values $\veps = \veps_n$, reasonably close to each other, so that we have to solve a series of eigenvalue problems:
\be 
H_n \Vc_n = \Vc_n E_n, \quad n=1,\dots, \mathscr{N}
\label{parametricEV}
\ee 
where $H_n \equiv H(\veps_n)$, $\Vc_n \equiv \Vc(\veps_n)$ and $E_n \equiv E(\veps_n)$. 
$\mathscr{N}$ is the number of different discrete values taken by the parameter.
In the molecule-field example introduced above, small values of $n$ will be typically associated with weak couplings (\textit{i.e.} diagonal-dominant matrices) and large values of $n$ will correspond to the strong-coupling case. 
We will not strive for all the eigenvalues and eigenvectors of Eq. \eqref{parametricEV} but rather try to obtain a subspace of $M$ states of interest among them. 

There are many popular ways of solving such block-eigenproblems. 
We would like to access large matrices so we focus on iterative methods in which storing the matrix can be avoided \cite{strang1986,saad2011}. Among them, we develop our reasoning about a particular projective method able to deal with non-hermitian matrices and very well-suited to the situation described above: the wave operator approach. This approach is usually split into two versions, a time-independent approach (the Bloch wave operator method), designed for finding the eigenstates and eigenvalues of a stationary hamiltonian \cite{reviewgeorges1}, and a time-dependent approach, used to facilitate the integration of the time-dependent Schr\"odinger equation \cite{reviewgeorges2,viennot2014}. In both cases, the wave operator transforms the original problem 
into a much simpler one, described by an effective $M\times M$ hamiltonian within an active subspace of reduced dimension $M \ll N$. 
In this paper we focus on the time-independent formulation,
where the effective hamiltonian given by the wave operator transformation gives $M$ exact eigenvalues and eigenvectors of the original problem after some iterative work. The wave operator can be explicitely calculated using an iterative process which only requires matrix-vector products. 
One possible choice is the Recursive Distorted Wave Approximation algorithm (RDWA) \cite{jolicard1987,reviewgeorges1}. 
The size of the working subspace does not grow during the iterations, and so does the memory cost. 
This is the main advantage of wave operator algorithms compared to other iterative methods. 

But things are not always so easy in practice. 
The main weakness of wave operator algorithms is related to their convergence radius. 
The effective, small-dimensional equivalent problem is defined \textit{via} the choice of an active (or model) subspace, usually chosen as the subspace spanned by some selected states vectors among those of the working basis set.
The selected vectors are expected to have important overlaps with the unknown eigenvectors. However, the model space should not be too far from the eigenspace to reach convergence 
(the quantum distance between two subspaces can be evaluated using the Fubini-Study distance \cite{viennot2007,viennot2014}).
When a coupling parameter progressively grows in the hamiltonian, the eigen-subspace to be calculated is gradually modified and moves away from the model space.
Typically a very good convergence is obtained when couplings between the selected active space and its complementary subspace are weak. Then the convergence usually gets gradually slower and slower with increasing couplings. 
When the distance between the model space and the eigenvectors reaches a certain threshold, the standard wave operator iteration or its variants fail to converge further. 
Similar instability problems can arise in the time-dependent formulation of the wave operator equations. A first attempt to handle this difficulty has been made in the context of nearly adiabatic quantum dynamics by defining a time-dependent adiabatic deformation of the active space \cite{viennot2014}. Such an adiabatic deformation implies the use of an efficient partial diagonalization method at each time step to be competitive.

In this article, we propose a time-independent wave operator iterative algorithm to partially diagonalize a parameter-dependent Hamiltonian, in which we introduce the concept of adaptative active subspace.
The general idea is to climb the parametric problem step by step, using the eigen-subspace associated with the previous hamiltonian $H_{n-1}$ as the active subspace for the wave operator associated with the next value of the parametric hamiltonian $H_{n}$.  
In that way it is expected that the quantum distance between the active subspace and the wanted eigenspace remains small all along the parameter iterations.
The active subspace just follows the eigen-subspace one step back. 
Although the idea seems simple, a practical implementation of it requires some efforts and approximations to avoid storing any $N \times N$ matrix and to limit the number of required matrix-vector products during the iterations. 
%
Note that other efficient diagonalization methods are based on the iterative construction of successive working bases\cite{scribano2008,garnier2016,lesko2019}. 
In those methods, the working space is enlarged and refined until the convergence of the eigenvalues for one given hamiltonian. 
Our point of view is a bit different since we deal with series of hamiltonians depending on some external physical parameter. The active subspace will stay fixed for a given hamiltonian and its dimensionality does not grow during the iterations, thanks to the properties of the wave operator algorithm. It is updated only between two successive values of the external parameter. 

In section \ref{fixedAS}, we summarise the main equations useful to find the standard Bloch wave operator using a fixed active space. 
The concept of an adaptative active subspace is developed in section \ref{adaptativeAS} and the corresponding iterative algorithm is presented. 
The main equations defining the adaptative wave operator are introduced in subsection \ref{AWOequations}
 and the internal working equations are derived in subsection \ref{internal}. 
We illustrate the capabilities of the method with numerical tests in sections \ref{numerical1} and  \ref{numerical2}. 
In section \ref{numerical1} we consider a model vibrational hamiltonian describing four linearly-coupled harmonic oscillators 
and we compare the results given by the adaptative wave operator method with the standard wave operator method and with a block-Davidson algorithm
\cite{davidson1975,liu1978,saad2011,ribeiro2005} widely used in molecular dynamics and quantum chemistry. 
In section \ref{numerical2} we examine the case of a diatomic molecular H$_2^+$ ion submitted to a strong laser field within the framework of  Floquet theory. 
In this case the Floquet hamiltonian matrix is non-hermitian and the spectrum features several exceptional points, rendering the challenge more interesting\cite{moiseyevNHQM}. 
We calculate maps of photodissociation resonance eigenvalues as a function of the laser wavelength and intensity.
Such maps can be very useful to design adiabatic control strategies in the context of molecular cooling\cite{lefebvre2012,leclerc2017}.

\section{Standard wave operator equations using a fixed active space \label{fixedAS}}

\subsection{The Bloch equation} 

Here we summarize the philosophy and the main equations of the Bloch wave operator theory using a fixed active space. 
More details can be found in the review articles by Killingbeck and Jolicard \cite{reviewgeorges1,reviewgeorges2}. 
All the matrices are expressed in some primitive basis set for the finite $N$-dimensional Hilbert space denoted $\{ \ket{u_i}, i=1,\dots,N \}$.
Let $S_0$ be the active subspace spanned by a collection of $M$ vectors selected among the $N$ vectors of the working basis set, with $M \ll N$. The basis is assumed to be re-ordered such that the selected vectors have indices running from $1$ to $M$:  $\{ \ket{u_i}, \; i=1,\dots,M\}$.
With this definition, matrices can be splitted into four blocks using projectors. 
Let $P_0$ be the projector on the active subspace $S_0$, 
\be 
P_0 = \sum_{i = 1}^M \ket{u_i} \langle u_i \vert =
\left(  
\begin{array}{c|ccc}
\text{I}_{M}	& 0	& \dots & 0 \\ 
\hline 
0      &  0  	&  \dots	&  0 \\ 
\vdots &   \vdots & \ddots & \vdots \\ 
0 	   & 0 		& \dots 	&  0 \\ 
\end{array} 
\right),
\label{projP0}
\ee 
where $I_M$ denotes the $M \times M$ identity matrix. 
The above matrix partitioning corresponds to the following block sizes:
\be 
\left(  
\begin{array}{c|c}
 M\times M & M\times(N-M) \\
 \hline 
 (N-M)\times M & (N-M)\times(N-M) \\
\end{array} 
\right).
\ee 
We will not use the complete matrix notation very often in the rest of the paper but every time a matrix is shown as blocks, this means that the above block sizes are assumed. 
The complementary subspace $S_0^{\dagger}$ corresponds to the projector $Q_0 = \text{I}_N - P_0$, i.e.
\be 
Q_0 = \sum_{i = M+1}^N \ket{u_i} \langle u_i \vert =
\left(  
\begin{array}{c|c}
  0	    &  0\\ 
  \hline 
0 	   &   \text{I}_{N-M} \\ 
\end{array} 
\right).
\label{projQ0}
\ee 
The Bloch wave operator $\Omega$ is designed to transform the original $N\times N$ hamiltonian $H$ into an effective $M\times M$ hamiltonian $H_{\text{eff}}$. The wave operator $\Omega$ has the form \cite{reviewgeorges1,reviewgeorges2}
\be 
\Omega = P_0 + Q_0 X P_0 = 
\left(  
\begin{array}{c|c}
I_M & 0 \\
 \hline 
X & 0 \\
\end{array} 
\right)
\label{blochwo}
\ee 
where $X=Q_0 X P_0$ is a block of size $(N-M)\times M$ as indicated above, 
and it satisfies the following non-linear equation:
\be 
H \Omega = \Omega H \Omega. 
\label{blocheq}
\ee 
The associated effective $M \times M$ hamiltonian is defined as
\be 
H_{\text{eff}} = P_0 H \Omega P_0.
\label{Heff} 
\ee 
If Eqs.\eqref{blochwo} and \eqref{blocheq} are satisfied, then the diagonalisation of $H_{\text{eff}}$ gives
\be 
H_{\text{eff} } T = T D
\label{eigenHeff}
\ee  
where $D$ is the $M\times M$ diagonal eigenvalue matrix and $T$ is the $M\times M$ matrix of eigenvectors of $H_{\text{eff}}$. 
The main interest of the wave operator approach is that 
the eigenvalues in $D$ are also {\emph{some of the exact eigenvalues}} of the original complete hamiltonian $H$ (previously denoted with symbol $E$). 
The wave operator $\Omega$ provides $M$ eigenvectors associated with the original problem, through the transformation 
\be 
V = \Omega T,  
\label{eigenvectorsH}
\ee 
where we have collected the eigenvectors of the original hamiltonian $H$ in an $N\times M$ rectangular matrix $V$,
\be
HV = VD.
\ee  

\subsection{Iterative solution}

In practice an iterative algorithm must be used to find $\Omega$ and $H_{\text{eff}}$ from the non-linear equations \eqref{blochwo}-\eqref{Heff}. 
Eq.~\eqref{blocheq} is reformulated as an equation in $X$, the non diagonal block of $\Omega$, which couples the active subspace to its complementary subspace (see Eq. \eqref{blochwo}), leading to 
\be 
X H_{\text{eff}} = Q_0 H P_0 + Q_0 H Q_0 \; X. 
\label{bloch2}
\ee
The first objective is to obtain a self-consistent equation for $X$ of the form $X = \mathcal{F}(X)$, from which an iterative process can be defined. 
One common approach consists in introducing a test diagonal matrix $H'$ restricted to the complementary space $S_0^{\dagger}$, i.e. $H'= Q_0 H' Q_0$ and $H'_{ij}=H'_{ii}\delta_{ij}$. The matrix $H'X$ is substracted from Eq.\eqref{bloch2} as a preconditioning step,
\be 
X H_{\text{eff}} - H'X = Q_0 H P_0 + Q_0 H Q_0 X - H'X. 
\label{bloch3}
\ee
Then we write $H_{\text{eff}}$ as its spectral representation, $H_{\text{eff}} = T D T^{-1}$, and multiply \eqref{bloch2} on the right by $T$ to obtain
\be 
(X T) D  - H' (X T) =  \left[ Q_0 H P_0  +  (Q_0 H Q_0 - H') X \right] T. 
\label{eqXT}
\ee 
$D$ and $H'$ are diagonal matrices in $S_0$ and $S_0^{\dagger}$, respectively, i.e.
\be
D =  
\left(  
\begin{array}{cc|c}
D_{11} {}_{\ddots} & 0  &  0 \dots \\
 0 & D_{MM} &  \\
 \hline 
0  &  &   0_{\ddots}  \\
\vdots  & &  \\
\end{array} 
\right)
\ee
and
\be 
H'=
\left(  
\begin{array}{c|cc}
^{\ddots}0 \; \;	& 0	& \dots  \\ 
\hline 
0      &  H'_{M+1,M+1} {}_{\ddots} 	&  0 \\ 
\vdots &   0 & H'_{NN} \\ 
\end{array} 
\right).
\ee 
Thus Eq. \eqref{eqXT} can be inverted to get 
\be 
X = \mathcal{I}(X) T^{-1}
\label{XFX}
\ee 
where the matrix $\mathcal{I}$ depends on $X$ and is defined by the elementwise expression:
\bea 
\mathcal{I}_{ij}(X) 
&=& \frac{ 
\left\{ \left[ Q_0 H P_0 + Q_0 (H-H') Q_0 X \right]  T  \right\} _{ij}
 }{ 
 D_{jj} - H'_{ii}  
 }, \nonumber  \\
&&
\left\{ 
\begin{array}{l}
 i  = (M+1),\dots,N \\ 
 j = 1,\dots,M.
\end{array} 
\right.
\label{Idef}
\eea
Note that matrices $D$ and $T$ which diagonalize $H_{\text{eff}}$ both depends on $X$ in Eq.~\eqref{Idef}. 
The practical iterative process is based on a Newton-Raphson argument \cite{durand1983} or equivalently on the application of the fixed point theorem \cite{theseDavid} to Eq. \eqref{XFX}, to make a sequence $X^{(p)}$ converging to $X$: 

\be
X^{(p+1)} = \mathcal{I}(X^{(p)}) [T^{(p)}]^{-1}. 
\label{staticit}
\ee
A convenient starting wave operator is $X^{(0)}=0$, or equivalently $\Omega^{(0)} = P_0$, or any avalaible approximation of $\Omega$ given by previous calculations.
The preconditioning matrix $H'$ can simply be made of the diagonal elements of $H$, $H'= \text{diag}(Q_0 H Q_0)$. This is similar to the preconditioner of the Davidson method \cite{davidson1975,saad2011}. Another choice is the Recursive Distorted Wave Approximation (RDWA), using  $H'= Q_0(1-X)H(1+X)Q_0$. 

The above integration procedure have distinctive features which render it very effective when compared to more conventional perturbative approaches. Working in a degenerate active space $S_0$ avoids divergences due to accidental degeneracies within this space. In the context of a varying hamiltonian introduced in section \ref{introduction}, the wave operator associated with a given value of the coupling parameters can be used as a starting point for the calculation of the wave operator associated with the next value of the couplings. 
However, the above strategy has a finite convergence radius in terms of the quantum Fubini-Study distance between the (unknown) eigen-subspace and the selected active space $S_0$; this can lead to numerical divergence, especially when sequences of calculations are intended to be done using parameter-dependent hamiltonians associated to strong-field coupling terms between quantum states of a molecule\cite{leclerc2017}. 
The divergence occurs when the Fubini-Study distance approaches the limit value $\pi/2$ which corresponds to an orthogonality between the eigen-subspace and the selected active space $S_0$ \cite{viennot2007,viennot2014,jolicard2016}. 
It is therefore necessary to develop a more powerful algorithm able to distord the active subspace and to make it follow more closely the unknown eigenspace.

\section{Wave operator algorithm for parameter-dependent hamiltonians using adaptative active spaces \label{adaptativeAS}}

\subsection{Parameter-dependent hamiltonian}

We now consider the parameter-dependent eigenvalue problem of Eq. \eqref{parametricEV},
as defined in section \ref{introduction}. 
No specific form is assumed for the hamiltonian (\textit{i.e.} we do not require it to be a sum-of-product), since we focus on the iterative algorithm itself which is problem-independent. As in any iterative method, the hamiltonian matrix elements does not need to be explicitely calculated or stored in memory. Only matrix-vector products are required, the efficiency of which will depend on the particular form of the hamiltonian. 

We assume that $H$ depends on one or several continuous parameters denoted collectively by $\veps$. The discretization of $\veps$ is $\veps_n$, with $n=0,\dots, \mathscr{N}$. 
In what follows the index $n$ will be used to label the different values of the coupling parameter whereas the index $p$ counts the iterations of the iterative process to find the wave operator at each value of $n$. 
We assume that the start value $n=0$ corresponds to $H \equiv H_0$ whose eigenvectors are already known and define the primitive working basis set,
\be 
H_0 \ket{u_i} = E_{0,i} \ket{u_i}, \; i=1,\dots,N. 
\ee 
The hamiltonian is gradually modified when the parameters collected in $\veps$ progressively change. 
The small difference between two successive hamiltonians is denoted by
\be 
 \Delta H_{n-1} = H_{n} - H_{n-1}. 
\ee 
The current hamiltonian at step $n$ is 
\be 
H_n = H_0 + \sum_{n=0}^{\mathscr{N}-1} \Delta H_{n}. 
\ee 
At each step, we aim at finding $M$ eigenvalues and eigenvectors of $H_n$,
assuming that $M$ eigenvalues and eigenvectors of $H_{n-1}$ are already known:
\be 
H_{n-1} \ket{ v_{n-1,j}} =  E_{n-1,j} \ket{v_{n-1,j}}, \; j=1,\dots,M,
\ee 
or, in matrix form
\be 
H_{n-1} \Vc_{n-1} = \Vc_{n-1} E_{n-1}
\label{evpeqHn1}
\ee 
where $[\Vc_{n-1}]_{ij}$ is the $i^{th}$ component of the $j^{th}$ eigenvector $\ket{v_{n-1,j}}$ expressed in the $\{\ket{u_i}\}$ basis set and $E_{n-1}$ is the diagonal matrix of eigenvalues. 
The eigenvectors are assumed to be normalized,
\be 
\Vc_{n-1}^{\dagger} \Vc_{n-1} = I_M. 
\label{normalizationV}
\ee
Note that if $H_n$ is non-hermitian (but still symmetric), the left eigenvectors are the complexe conjugate of the right eigenvectors and we can adopt the c-product normalization convention \cite{moiseyevNHQM}, replacing 
$\Vc_{n-1}^{\dagger}$ 
with 
$\Vc_{n-1}^{t}$ in Eq.~\eqref{normalizationV} and subsequent equations. 

Knowing the $M$ eigenpairs of equation \eqref{evpeqHn1} associated with the previous value of the hamiltonian parameters, we wish to calculate $M$ eigenpairs of the current hamiltonian $H_n$, in matrix form
\be 
H_{n} \Vc_{n} = \Vc_{n} E_{n}. 
\label{evpeqHn}
\ee 
The purpose of the following subsections is to show that an adaptative wave operator algorithm can take into account the information associated with the previous value of the  parameters (eigenpairs $E_{n-1},\Vc_{n-1}$) to facilitate the search of the next eigenpairs ($E_{n},\Vc_{n}$) of a slightly different hamiltonian. 
The algorithm gives an updated effective hamiltonian matrix for each new value of the coupling parameters, whose eigenvalues are also exact eigenvalues of $H_n$.

\subsection{Wave operator equations using adaptative projectors \label{AWOequations}}

We wish to solve Eq. \eqref{evpeqHn} with $H_n = H_{n-1} + \Delta H_{n-1}$. 
At step $n$, the active subspace $S_{n}$ is defined as the subspace spanned by the $M$ known eigenvectors of $H_{n-1}$. The associated projector is
\be 
P_n = \sum_{j = 1} ^ M \ket{ v_{n-1,j} } \langle v_{n-1,j} \vert,
\ee 
or in matrix form
\be 
P_n = \Vc_{n-1} \; \Vc_{n-1}^{\dagger}.
\label{projPn}
\ee 
The projector on the complementary space is defined as 
\be 
Q_n = I_N - P_n
\label{projQn}
\ee 
At this stage there are two options. The first one (i) consists in a change of the working basis set to recover the usual equations for the Bloch wave operator. In this case the basis transformation would be non-orthogonal and imply $N \times N$ transformation matrices. 
The second option (ii) is based on working equations expressed directly in the primitive basis set $\{ \ket{u_j} \}$, using projectors such as those defined in Eqs.~(\ref{projPn}) and (\ref{projQn}), when needed in the calculations. In what follows we focus on the second option. 
Note that the derivation below remains valid to calculate the wave operator with any non-canonical projector, even if it has not been obtained in the context of a parameter-dependent hamiltonian but comes, for example, from some approximate solution. 

The Bloch wave operator equations are
\be 
H_n \Omega_n = \Omega_n H_n \Omega_n
\label{BWOn}
\ee 
with
\be 
\Omega_n = P_n + Q_n X_n P_n,
\ee
and the associated effective hamiltonian reads
\be 
H_{\text{eff},n} = P_n H_n \Omega_n P_n.
\label{eqHeffn}
\ee 
Here $P_n$ and $Q_n$ also cause $N \times N$ matrices. Yet it is possible to avoid this by using decomposition \eqref{projPn} in which $\Vc_{n-1}$ is a tractable matrix with only $M$ columns. 
As shown in Appendix \ref{AppCalc}, Eqs.~\eqref{BWOn}-\eqref{eqHeffn} can be recasted into the following more compact formulae:
\be 
Y_n \mathcal{M}_{\text{eff},n} = Q_n H_n \Vc_{n-1} + Q_n H_n Q_n \; Y_n,
\label{BWOneq1}
\ee
where the unknowns are
\be 
Y_n  \equiv Q_n X_n \Vc_{n-1},
\label{Yn}
\ee 
{\textit{i.e.}} the part of the wave operator which couples the active subspace $S_n$ to its complementary subspace $S_n^{\dagger}$,
and 
\be 
\mathcal{M}_{\text{eff},n} \equiv 
\Vc_{n-1}^{\dagger} H_n \Omega_n \Vc_{n-1},
\label{Meffn}
\ee 
which is the matrix representation of $H_{\text{eff},n}$ expressed in the $\Vc_{n-1}$ basis. 
Note that the rectangular matrix $Y_n$ satisfies the projective property
\be 
Y_n = Q_n Y_n.
\label{eqprojYn}
\ee 
If we also define the closed wave operator $\tilde{\Omega}_n$ as 
\be 
\tilde{\Omega}_n \equiv \Omega_n \Vc_{n-1},
\label{Omegatilde}
\ee 
we can note that 
\begin{eqnarray}
\tilde{\Omega}_n &=& ( \Vc_{n-1} \Vc_{n-1}^{\dagger} + Q_n X_n \Vc_{n-1} \Vc_{n-1}^{\dagger} ) \Vc_{n-1} \nonumber \\
 &=& \Vc_{n-1} + Y_n .
 \label{Omegatilde2}
\end{eqnarray}
and 
\be 
\mathcal{M}_{\text{eff},n} = \Vc_{n-1}^{\dagger} H_n \tilde{\Omega}_n.
\label{Meff2}
\ee 
The above equations \eqref{BWOneq1} to \eqref{Omegatilde} only imply small or rectangular matrices.
$Y_n$ and $\tilde{\Omega}_n$ are rectangular $N\times M$ matrices
(whereas $X_n$ and $\Omega_n$ were originaly full $N\times N$ matrices in the primitive basis set), 
and 
$\mathcal{M}_{\text{eff},n}$ is a small $M\times M$ square matrix. 
Matrices $H_n$ and $\Vc_{n-1}$ are already known from the previous calculation at the parameter step $(n-1)$.
The left multiplication by $Q_n$ can be evaluated using Eq.~\eqref{projQn}.

The difficulty of the calculation is hidden in Eq.~\eqref{BWOneq1}, the structure of which is similar to the Bloch equation \eqref{bloch2} in the fixed active space formulation of section \ref{fixedAS}. 
The main differences are that \\
(i) $(Q_n)Y_n$ has $M$ full columns, while $X$ in Eq.~\eqref{bloch2} has zeros in its first $M \times M$ square; \\
(ii) $\Vc_{n-1}$ has $M$ full columns, while $P_0$ in Eq.~\eqref{projP0} is a simple canonical projector (i.e. a $M\times M$ identity filled with zeros on the remaining components); \\
(iii) $Q_n H_n$ replaces $H$; \\
(iv) The effective hamiltonian $H_{\text{eff},n}$ is replaced by $\mathcal{M}_{\text{eff},n}$. \\
Using Eqs.~\eqref{eqHeffn} and \eqref{Meffn}, 
the effective hamiltonian associated to $H_n$ can also be written as
\begin{eqnarray}
H_{\text{eff},n} &=& \Vc_{n-1} \; \mathcal{M}_{\text{eff},n} \; \Vc_{n-1}^{\dagger} \nonumber \\
&=&  \Vc_{n-1} ( \Tc_n  D_n \Tc_n^{-1} ) \Vc_{n-1}^{\dagger} \label{diagMeff} \\ 
&=& T_n D_n T_n^{-1} \quad \text{  with  } T_n = \Vc_{n-1} \Tc_n \label{diagHeff}
\end{eqnarray}
where $\mathcal{T}_n$ is the $M\times M$ matrix which diagonalizes $\mathcal{M}_{\text{eff},n}$, with $D_n$ the corresponding diagonal matrix of eigenvalues. 
In Eq.~\eqref{diagHeff}, $H_{\text{eff},n}$ is a $N \times N$ matrix and
$T_n$ is a $N\times M$ matrix representing $M$ eigenvectors of $H_{\text{eff},n}$ in the primitive basis set. 
But $H_{\text{eff},n}$ and $\mathcal{M}_{\text{eff},n}$ share the same eigenvalues and it is obviously much easier to diagonalize $\mathcal{M}_{\text{eff},n}$ (which is $M \times M$) than $H_{\text{eff},n}$ (which is $N \times N$). 
The diagonalization of the small matrix $\mathcal{M}_{\text{eff},n}$ gives $M$ exact eigenvalues of the original hamiltonian $H_n$. 
The new eigenvectors $\Vc_n$ of $H_n$ can be simply recovered from Eq.~\eqref{eigenvectorsH}, adapted to the present situation,
\begin{eqnarray}
\Vc_n 	&=& \Omega_n P_n T_n  \nonumber \\
		&=& (\Omega_n \Vc_{n-1}) (\Vc_{n-1}^{\dagger} T_n) \nonumber \\
		&=& \tilde{\Omega}_n \mathcal{T}_n 
		\label{Vcnnouv}
\end{eqnarray}
where no $N \times N$ matrix is required since $\tilde{\Omega}_n$ is $N \times M$ and $\mathcal{T}_n$ is $M \times M$. 
Eq.~\eqref{Vcnnouv} shows that the eigenvectors can be obtained directly from the diagonalization of $\mathcal{M}_{\text{eff},n}$, there is no need to explicitely compute the product $T_n = \Vc_{n-1} \mathcal{T}_n$. 

To summarize, operations should be done in the following order (this is the main algorithm):  \\
\begin{enumerate}
	\item Initialize: at the very first iteration of the coupling parameter values, $\Vc_{n=0}$ is a canonical basis of dimension $M$ associated with the zero-order active subspace, $\tilde{\Omega}_{n=0} = \Vc_{n=0}$, and $\mathcal{M}_{\text{eff},n}$ is filled in with the corresponding diagonal elements of $H_{n=0}$. 
	\item Find a way to solve the main Eq.~\eqref{BWOneq1} for $Y_n$ and $\mathcal{M}_{\text{eff},n}$ with the current hamiltonian $H_n$. For this we will develop an internal iterative algorithm in the following subsection \ref{internal}; \label{mainstep}
	\item Compute the wave operator $\tilde{\Omega}_n = \Vc_{n-1} + Y_n$ (Eq.~\eqref{Omegatilde2});
	\item Diagonalize $\mathcal{M}_{\text{eff},n}$ using a direct method to find $\mathcal{T}_n$ and $D_n$ (Eq.~\eqref{diagMeff}). $D_n$ contains $M$ exact eigenvalues of the original hamiltonian $H_n$;

	\item Compute the eigenvectors $\Vc_n=\tilde{\Omega}_n \Tc_n$ (Eq.~\eqref{Vcnnouv}).
	\item Iterate over the parameter index, $n \leftarrow n+1$:
   	\begin{enumerate}
    	\item update the hamiltonian $H_{n} \leftarrow H_{n+1}$ with the new values of the coupling parameter $\veps_{n+1}$
    	\item update the projector $P_n \leftarrow P_{n+1} = \Vc_{n} \; \Vc_{n}^{\dagger}$
    	\item go back to step 2.
	\end{enumerate}
\end{enumerate} 

Step \ref{mainstep} in the above algorithm have to be clarified and deserves special attention. The objective of the following subsection is to propose an iterative algorithm to solve Eqs.~\eqref{BWOneq1} to \eqref{Meffn}.

\subsection{Internal iterative algorithm \label{internal}} 

In addition to the external loop over the parameter value (labeled using index $n$), an internal iterative algorithm is proposed here (with iterations labeled using index $p$) to find the wave operator $\tilde{\Omega}_n$, via the search of matrices $Y_n$ and $\mathcal{M}_{\text{eff},n}$, in the same spirit of what have been summarized in section \ref{fixedAS} for the case of a fixed active space. 
We start by inserting $\mathcal{M}_{\text{eff},n} = \Tc_n  D_n \Tc_n^{-1} $ (see Eq.~\eqref{diagMeff}) into Eq.~\eqref{BWOneq1} to get
\be 
Q_n Y_n \Tc_n D_n = Q_n H_n Y_n \Tc_n + Q_n H_n \Vc_{n-1} \Tc_n.
\ee  
We note that $H_n = H_{n-1} + \Delta H_{n-1}$ and make use of Eq.~\eqref{evpeqHn1} and of the identity $Q_n \Vc_{n-1} = 0$, this leads to
\be 
Q_n Y_n \Tc_n D_n = Q_n H_n Y_n \Tc_n + Q_n \Delta H_{n-1} \Vc_{n-1} \Tc_n.
\label{eqYn}
\ee 
Assuming that $Y_n^{(p-1)}$ is an imperfect solution to Eq.~\eqref{eqYn}, we ask for a refined solution
\be 
Y_n^{(p)} = Y_n^{(p-1)} + \Delta Y_n^{(p-1)}
\label{updateY}
\ee 
which satisfies \eqref{eqYn}. 
$\Tc_n^{(p-1)}$ and $D_n^{(p-1)}$ are the matrices which diagonalize the current effective matrix $\mathcal{M}_{\text{eff},n}^{(p-1)}$ associated with $Y_n^{(p-1)}$, i.e. $\mathcal{M}_{\text{eff},n}^{(p-1)} = \Vc_{n-1}^{\dagger} H_n ( \Vc_{n-1} + Y_n^{(p-1)} )$, see Eqs.~\eqref{Omegatilde2} and \eqref{Meff2}. 
Thus Eq.~\eqref{eqYn} becomes
\bea
&Q_n  & \left[ Y_n^{(p-1)} + \Delta Y_n^{(p-1)} \right] \Tc_n^{(p-1)} D_n^{(p-1)}  \nonumber \\
&=& 
Q_n H_n \left[ Y_n^{(p-1)} + \Delta Y_n^{(p-1)} \right] \Tc_n^{(p-1)} \nonumber \\
&& + Q_n \Delta H_{n-1} \Vc_{n-1} \Tc_n^{(p-1)}.
\eea
From now on, the superscripts $(p-1)$ on $\Tc_n$, $D_n$, $\Delta Y$ and subscripts $(n-1)$ on $\Delta H$ are temporarily ignored for simplicity. The unknown increment $\Delta Y$ satisfies the following non-linear equation:
\be
Q_n \,\, \Delta Y \,\,  \Tc_n D_n - Q_n H_n \,\, \Delta Y \,\, \Tc_n 
=
\mathcal{F}(Y_n^{(p-1)})
\label{eqDeltaY}
\ee 
where
\bea
 \mathcal{F}(Y_n^{(p-1)}) & \equiv &
Q_n H_n Y_n^{(p-1)}\Tc_n  \nonumber \\
& -&Q_n Y_n^{(p-1)}  \Tc_n D_n 
+ Q_n \Delta H \Vc_{n-1} \Tc_n  
\label{eqrighthandside}
\eea
is a $N\times M$ matrix.      
In the left-hand-side of Eq.~\eqref{eqDeltaY}, $D_n$ is diagonal but $Q_nH_n$ is not. 
Eq.~\eqref{eqDeltaY} cannot be easily inverted to extract the optimal increment $\Delta Y$ needed to refine $Y_n^{(p-1)}$. 

We propose to build a second, sub-level of iterations to find the increment $\Delta Y$. 
We consider $Y_n^{(p-1)}$ temporarily fixed  
and we seek the rectangular $N\times M$ matrix $Z$ defined as the product
\be 
Z = \Delta Y \Tc_n.
\ee
Eq.~\eqref{eqDeltaY} can be written in terms of $Z$ as
\be 
Z \, D_n - Q_n H_n \, Z = \mathcal{F}(Y_n^{(p-1)}) .
\label{eqZ}
\ee 
Eq.~\eqref{eqZ} can be solved iteratively
using a fixed-point argument as shown in Appendix \ref{AppZ}. 
Starting with a zero $N \times M$ matrix for $Z^{(0)}$, $Z$ can be found by building the sequence
\begin{widetext}
\be 
Z_{ij}^{(\ell)} = 
\frac{
\left[ \mathcal{F}(Y_n^{(p-1)}) + (Q_n H_n - H') Z^{(\ell-1)} - Z^{(\ell-1)}(D_n-D')  \right]_{ij}
}{
D'_{jj} - H'_{ii}
}, \quad \ell = 1,\dots,\mathscr{L}
\label{eqZijiteration}
\ee 
\end{widetext}
for $i=1,\dots, N$ and $j=1,\dots,M$.
In Eq. \eqref{eqZijiteration}, $D'$ and $H'$ are two arbitrary diagonal matrices of respective sizes $M\times M$ and $N\times N$.

In practice, we have to ensure that the projective property of Eq.~\eqref{eqprojYn} remains satisfied along the internal iterations. Eq.~\eqref{eqZijiteration} can jeopardize this projective property. 
It is thus necessary to explicitely apply the projector onto the complementary subspace to ensure the validity of this property between two successive iterations:
\be 
Z^{(\ell)} \leftarrow Q_n Z^{(\ell)}.
\label{eqprojZ}
\ee
It is not necessary to wait for a complete convergence of Eq.~\eqref{eqZijiteration} because this nested iteration concerns an increment of the main iterative process defined in Eq.~\eqref{updateY} and \eqref{eqDeltaY}. 
Assume that after $\mathscr{L}$ iterations a reasonable level of convergence is reached, the optimal increment $\Delta Y^{(p-1)}$ for Eq.~\eqref{updateY} is finally calculated as
\be 
\Delta Y_n^{(p-1)} =  Z^{(\ell=\mathscr{L})} \left[\Tc_n^{(p-1)} \right]^{-1}.
\label{calculDeltaY}
\ee 
To ensure the convergence of the above nested iterative procedure, the $D'$ and $H'$ preconditioning matrices should be chosen reasonably close to $D_n$ and $Q_n H_n$, respectively, as explained in Appendix \ref{appconvergence}.
A simple choice is to use the diagonal values of $H_n$: we use the first $M$ values to fill in the $D'$ diagonal and we use $M$ zeros followed by the remaining $(N-M)$ diagonal values of $H_n$ to fill in $H'$:
\bea
D' &=& 
\text{diag}(P_0 H_n P_0),  \nonumber \\
\quad 
H' &=& \text{diag}(Q_0  H_n Q_0)  = 
\left(  
\begin{array}{c|c}
 0_{M\times M} & 0 \\
 \hline 
 0 & \text{diag}(H_n) \\
\end{array} 
\right).
\label{eqprecond}
\eea 
Better choices are possible for this preconditioning step. The $D'$ diagonal matrix may be improved by using the eigenvalues of $H_{n-1}$ obtained at the previous coupling iteration,
\be
D' = D_{n-1}.
\label{betterDprime} 
\ee
The preconditioning matrix $H'$ may also be improved. 
Appendix \ref{appprecond} gives all the working equations in the case of a non-diagonal preconditioner $H'$ defined in a subspace of intermediate dimension $R$ such that $M < R \ll N$.
The main idea is to use the block matrix representation of $Q_n H_n Q_n$ within this intermediate subspace leading to an improved radius of convergence in the numerical applications.

There are finally three indices in the algorithm: index $n$ associated with the parameter values in the hamiltonian $H_n$; index $p$ which labels the successive approximations of the wave operator $\tilde{\Omega}_n^{(p)}$ \textit{via} the matrix $Y_n^{(p)}$; index $\ell$ corresponding to the nested working iteration used to find the increment $\Delta Y_n^{p-1}$ \textit{via } the search of $Z^{(\ell)}$. 
Most of the numerical work is done in the most internal iteration over $\ell$. 

Finally we give a sketch of the internal part of the algorithm, for a given value of the parameter index $n$ and with an active space projector defined by Eq.~\eqref{projPn}.
The algorithm below should be done as step \ref{mainstep} of the main algorithm given at the end of section \ref{AWOequations}:

\begin{enumerate}
\item Initialization:
	\begin{enumerate}
	\item Start with $Y_n^{(p=0)}$ a zero $(N \times M)$ matrix, 
	\item Set $\tilde{\Omega}_n^{(p=0)} = \Vc_{n-1} $;
	\item Calculate  $\mathcal{M}_{\text{eff},n}^{(p=0)} = \Vc_{n-1}^{\dagger} H_n \Vc_{n-1}$ and diagonalize it,
	 i.e. $\mathcal{M}_{\text{eff},n}^{(0)} \Tc_n^{(0)} = \Tc_n^{(0)} D_n^{(0)}$;
	\end{enumerate}
\item Calculate $\Delta Y^{(p-1)}$: \label{step1}
    \begin{enumerate}
      \item Calculate or update $\mathcal{F}(Y_n^{(p-1)})$ using Eq.~\eqref{eqrighthandside}; \label{stepF}
      \item Start with $Z^{(0)}$ an $(N \times M)$ zero matrix;
      \item Iterate Eq.~\eqref{eqZijiteration} for $Z^{(\ell)}$ using $H'$ and $D'$ defined in Eq.~\eqref{eqprecond}. Stop if $\frac{\Vert Z^{(\ell)}-Z^{(\ell-1)} \Vert}{ \Vert Z^{(\ell)} \Vert}$ is lower than some reasonable threshold or if $\ell = \mathscr{L}$; \label{stepZ}
      \item Calculate $\Delta Y^{(p-1)}$ using Eq.~\eqref{calculDeltaY};
    \end{enumerate}
\item Update $Y_n^{(p)} = Y_n^{(p-1)} + \Delta Y^{(p-1)}$, Eq.~\eqref{updateY};
\item Update the wave operator $\tilde{\Omega}_n^{(p)} = \Vc_{n-1} + Y_n^{(p)}$, Eq.~\eqref{Omegatilde2};
\item Update the effective hamiltonian $\mathcal{M}_{\text{eff},n}^{(p)} = \Vc_{n-1}^{\dagger} H_n \tilde{\Omega}_n^{(p)}$; \label{stepM}
\item Solve the eigenproblem $\mathcal{M}_{\text{eff},n}^{(p)} \Tc_n^{(p)} = \Tc_n^{(p)} D_n^{(p)}$ and calculate the current approximation of the eigenvectors (see section \ref{AWOequations}, Eq.~\eqref{Vcnnouv}); \label{directdiag}
\item Check the convergence (see subsection \ref{secconv}). If not reached, iterate $(p \leftarrow p+1)$ and go back to step \ref{step1}.
\end{enumerate}
The main difference with equations \eqref{Idef} and \eqref{staticit} in the static algorithm of section \ref{fixedAS} is that the iterative working equations of the adaptative algorithm concern a corrective term to $Y_n$, not $Y_n$ itself.

\subsection{Convergence criteria \label{secconv}}

Several criteria can be used to monitor the convergence. 
We can calculate the following self-consistent residual of Eq.~\eqref{BWOneq1} for the wave operator $\tilde{\Omega}$,
\be 
\delta_1 ^{(p)}= 
\frac{ 
\Vert Q_n H_n \tilde{\Omega}_n^{(p)} - Y_n^{(p)} \mathcal{M}_{\text{eff},n}^{(p)} \Vert_F 
}{ 
\Vert \tilde{\Omega}_n^{(p)} \Vert _F
}
\label{eqdelta1}
\ee
where $\Vert \cdot \Vert_F $ denotes the Frobenius norm. 
We can also check the traditional norm of the residual for the original eigenproblem,
\be 
\delta_2 ^{(p)} =
\frac{ 
\Vert H_n \Vc_n^{(p)} - \Vc_n^{(p)} D_n^{(p)} \Vert_F
}{
\Vert \Vc_n^{(p)} \Vert_F
}.
\label{eqdelta2}
\ee 
This global residual takes into account all $M$ eigenvectors of the block, each on the same footing. 
Since some eigenvalues may converge faster than others, we can also restrict the calculation of $\delta_2 ^{(p)}$ to some particular column(s). 
In practice, both residuals defined above are almost equivalent and the calculation is stopped when either $\delta_1 ^{(p)}$ and $\delta_2 ^{(p)}$ becomes lower than some convergence threshold $\delta_{\text{stop}} $. 

Note that in the internal algorithm, a stopping criterion is also needed for the most internal loop of Eq.~\eqref{eqZijiteration} over index $\ell$, used to find the optimal increment.
We choose to monitor the following relative residual: 
\be 
\delta_{\text{int}} ^{(\ell)} =
\frac{ 
\Vert 
Z^{(\ell)} - Z^{(\ell-1)}
\Vert_F
}{
\Vert 
Z^{(\ell)}
\Vert_F
}.
\label{eq_res_int}
\ee
The internal loop \eqref{eqZijiteration} is stopped if $\delta_{\text{int}} ^{(\ell)} < \delta_{\text{stop}^{\text{int}}}$.

\subsection{Numerical cost} 

The numerical cost is dominated by the matrix-vector products necessary to obtain 
the numerator of Eq.~\eqref{eqZijiteration}. 
We recall that $N$ and $M$ denotes the dimension of the total Hilbert space and the active subspace, respectively. 
Products of the form $(H_n Y_n)$, $(Q_n Y_n)$ and $(Q_n H_n Y_n)$ are needed in Eq.~\eqref{eqrighthandside} of step \ref{stepF} (section \ref{internal}).  
The dominant calculation is clearly the matrix product $H_n Y_n$ which costs $\mathcal{O}(MN^2)$. 
During step \ref{stepZ} (section \ref{internal}), a matrix product $H_n Z^{(\ell)}$ is also needed for each iteration over index $\ell$ to update the numerator of Eq.~\eqref{eqZijiteration} with the same cost $\mathcal{O}(MN^2)$. 
The same matrix vector product $H_n Y_n^{(p)}$ is needed to calculate both $\mathcal{M}_{\text{eff},n}^{(p)}$ in step \ref{stepM} and $\mathcal{F}(Y_n^{(p)})$ used during the next $(p+1)^{\text{th}}$ iteration, it does not need to be computed twice. 
The projections on $Q_n$ such as $Q_n Z^{(\ell)}$ or $Q_n(H_n Y_n)$ are always made using the identity $Q_n = I_N - \Vc_{n-1} \Vc_{n-1}^{\dagger}$ and thus 
costs $\mathcal{O}(2 M^2 N)$. 
Right-multiplications by $\Tc_n$ cost $\mathcal{O}(M^2 N)$ more. 
The third term $Q_n \Delta H \Vc_{n-1}\Tc_n $ in Eq.~\eqref{eqrighthandside} costs $\mathcal{O}( M N^2 + M^2 N)$ but the matrix vector product 
$Q_n \Delta H \Vc_{n-1}$ can be calculated only once for each new value of the parameters $\veps_n$. 
The direct diagonalization in step \ref{directdiag} costs $\mathcal{O}(M^3)$ which is negligible if $M \ll N$ as stated here. 
In conclusion, the overall cost of the total algorithm to find $M$ eigenpairs of a $N\times N$ parametric hamiltonian scales as
\be 
\mathcal{O}( \mathscr{N} \mathscr{P} \mathscr{L} MN^2)
\ee
where $\mathscr{N}$ is the number of different values taken by the parameters in the hamiltonian, $\mathscr{P}$ and $\mathscr{L}$ are typical values of the maximum number of iterations needed to reach convergence in the internal algorithm of section \ref{internal}.

As in any other iterative method, the hamiltonian matrix does not need to be fully stored in memory. The memory cost is the cost of vectors. It is related to rectangular matrices to store the wave operator, its increment and eigenvectors so it scales as $\mathcal{O}(NM)$. Contrary to Krylov subspace methods\cite{saad2011} such as Lanczos or Davidson algorithms, the memory cost remains stable and does not grow during the iterations. This advantage could become of particular interest in multidimensional applications where very large vectors or tensors have to be stored in memory.


\section{Application to a model parameter-dependent vibrational hamiltonian \label{numerical1}}

\subsection{Model hamiltonian and theoretical eigenvalues} 

We first apply the adaptative wave operator algorithm to an ensemble of bilinearly coupled harmonic oscillators with varying coupling. 
In this case the hamiltonian is hermitian. The model has the advantage that the exact energy levels are analytically known. 
The hamiltonian is 
\be 
H (q_1,\dots,q_D) =  \sum_{j=1}^D \frac{\omega_j}{2} \left( p_j^2 + q_j^2 \right) +
\veps \;
\sum_{\substack{i,j=1 \\ i>j}}^D \alpha_{ij} q_i q_j
\label{coupledhamiltonian}
\end{equation}
with $q_j$ the $j^{\text{th}}$ normal coordinate and $p_j = - \imath \frac{\partial}{\partial q_j}$.
For simplicity we set a single coupling constant, 
\be 
\alpha_{ij} =1  \quad \forall i,j.
\label{eqsinglecoupling}
\ee 
The coupling parameter $\veps$
 will be varied from $0$ to some maximum value $\veps_{\text{max}} $.
In most of the calculations below we choose to work with $D=4$ oscillators. 
The primitive basis set is build from $m_j$ harmonic oscillator functions for each coordinate, $\prod_{j=1}^D \theta_{i_j}^j (q_j)$
with $ \theta_{i_j}^j (q_j) $ the eigenfunctions of $\frac{\omega_j}{2} \left( p_j^2 + q_j^2 \right)$, $i_j = 1, \dots, m_j$. 
The complete direct product basis set is made of $N = \prod_{j=1}^D m_j$ functions. 
The advantage of this simple model hamiltonian is that exact eigenvalues of $H$ can be obtained by transforming to normal coordinates. They are given by
\begin{equation}
E_{k_1, \dots, k_D} = \sum_{j=1}^D \nu_j \left( \frac{1}{2} + k_j \right)\text{, with } k_j =0,1,\cdots 
\label{theoretical_spectrum}
\end{equation}
where the transformed frequencies $\nu_j$ are square roots of the eigenvalues of a matrix ${A}$ 
whose  elements are $A_{ii} = \omega_i^2$ and 
$A_{ij} = \veps \; \alpha_{ij} \sqrt{\omega_i} \sqrt{\omega_j}$. 
Since we limit our model system to four dimensions, we do not need to introduce any advanced strategies to reduce the basis size such as pruning or contractions. 
Our present purpose is not to tackle the dimensionality problem but to focus on the numerical behavior of the iterative algorithm in the context of parameter-dependent matrices. 
The model is simple but it is an interesting test for the adaptative wave operator algorithm. 
In what follows the coupling parameter $\veps$ is varied from $0$ to $0.2$ with $\mathscr{N} = 100$ intermediate steps. 
Table \ref{tab_common_parametres_coupled_osc} summarizes the main numerical parameters common to all the calculations of this section. 
The active subspace has dimension $M=20$ and is initially set as the lowest-lying uncoupled states. 
\begin{table}[ht]
\caption{Common numerical parameters for the $4D-$coupled oscillator calculations.}
\begin{tabular}{lr}
\hline
\hline
Number of $1D$ basis functions  	& $m_j=8$  \\
Total Hilbert space dimension 		& $N=4096$	 \\
Active subspace dimension 			& $M=20$ \\
Frequencies (arb. units)  &  $\omega_1 = \sqrt{2}$,  $\omega_2 = \sqrt{3}$ \\ 
& $\omega_3 = \sqrt{5}$,  $\omega_4 = \sqrt{7}$ \\ 
Coupling parameter step  & $\Delta \veps = 0.002$  \\
Number of parameter steps  & $\mathscr{N} = 100$  \\
\hline
\hline
\end{tabular}
\label{tab_common_parametres_coupled_osc}
\end{table}

\subsection{Numerical eigenvalues and convergence \label{sub_internal}}

We focus on three calculations along the coupling parameter trajectory: 
(i) a weak coupling case, with $\veps = 0.020$, corresponding to the parameter iteration $n=10$;
(ii) a moderate coupling case, with $\veps = 0.080$ (iteration $n=40$);
and
(iii) a stronger coupling case, with $\veps = 0.150$ (iteration $n=75$). 
Each calculation makes use of the previous eigenvectors to define the projector onto the active subspace. For example the moderate coupling case $\veps = 0.080$ uses the results associated with $\veps = 0.078$ as the active subspace.

\begin{table*}[ht]
\caption{Eigenvalues of the 4D-coupled oscillator model obtained using the adaptative wave operator algorithm. 
The first column shows the global eigenvalue label, the second column contains the theoretical eigenvalues (exactly known here), the third and fourth columns show the numerical results given by calculations (a) and (b) as defined in table \ref{tab_internal_par_effect}. For each eigenvalue the last correct digit is written in bold type.
We show the results at two different values of the coupling parameter (moderate or strong). }
\begin{tabular}{cccc}
\hline	
\hline	
Global label & 	Theoretical eigenvalues	&	$\quad$ Calculation (a) $\quad$ & $\quad$	Calculation (b)	$\quad$\\
\hline	
 \multicolumn{4}{c}{ 	\textit{Coupling parameter}		 $\veps=	0.08$	}		\\
\hline	
1	 & 	4.01169503098439	&	4.01169503098{\bf{4}}40	&	4.0116950309843{\bf{9}}	\\
2	&	5.41754357042936	&	5.417543570429{\bf{3}}7	&	5.417543570429{\bf{3}}7	\\
3	 & 	5.74179010128007	&	5.741790101280{\bf{0}}9	&	5.74179010128{\bf{0}}10	\\
4	&	6.24709816663631	&	6.247098166636{\bf{3}}3	&	6.247098166636{\bf{3}}3	\\
5	 & 	6.66373834756062	&	6.6637383475606{\bf{2}}	&	6.6637383475606{\bf{2}}	\\
6	&	6.82339210987433	&	6.8233921098743{\bf{3}}	&	6.823392109874{\bf{3}}4	\\
\vdots & \vdots & \vdots & \vdots \\
17	 & 	8.89914148321253	&	8.899141483212{\bf{5}}6	&	8.899141483212{\bf{5}}6	\\
18	&	9.05879524552624	&	9.058795245526{\bf{2}}7	&	9.058795245526{\bf{2}}6	\\
19	 & 	9.20198024187143	&	9.20198024187{\bf{1}}59	&	9.20198024187{\bf{1}}57	\\
20	&	9.31578166413684	&	9.315781664136{\bf{8}}5	&	9.315781664136{\bf{8}}7	\\
\hline							
 \multicolumn{4}{c}{ 	\textit{Coupling parameter}	 $\veps=	0.15$	}		\\
\hline							
1	 & 	4.00602786977868	&	Fails to converge 	&	4.006027869778{\bf{6}}9	\\
2	&	5.39412280725013	&		&	5.394122807250{\bf{1}}5	\\
3	 & 	5.72955426987126	&		&	5.729554269871{\bf{2}}8	\\
4	&	6.23770385197413	&		&	6.2377038519741{\bf{3}}	\\
5	 & 	6.67478628957654	&		&	6.674786289576{\bf{5}}7	\\
6	&	6.78221774472158	&		&	6.78221774472{\bf{1}}91	\\
\vdots & \vdots &   & \vdots \\
17	 & 	8.90646227177199	&		&	8.9064622717719{\bf{9}}	\\
18	&	9.01389372691704	&		&	9.0138937269{\bf{1}}884	\\
19	 & 	9.17660707005643	&		&	9.17660707{\bf{0}}11161	\\
20	&	9.34354470937440	&		&	9.343544709374{\bf{4}}3	\\
\hline 							
\hline 							
\end{tabular}
\label{tab_eigenvalues_coupledOH}
\end{table*}
We first compare the results given by two calculations using either (a) the diagonal preconditioner, or (b) the block-diagonal preconditioner described in appendix \ref{appprecond}, with the exact theoretical results. The internal stopping criterion for the nested iterations (see Eq.~\eqref{eq_res_int}) is set to $\delta_{\text{stop}^{\text{int}}} = 10^{-1}$. 
The global convergence criteria was chosen as $\delta_{1/2} < \delta_{\text{stop}} = 10^{-12}$. 
The size of the intermediate subspace is set to $R= 2 M = 40$
and it is made of the lowest energy uncoupled states. 
Table \ref{tab_eigenvalues_coupledOH} shows some of the eigenvalues obtained with the adaptative wave operator algorithm at moderate and strong coupling values. 
Calculation (a) and (b) converge well for all the wanted eigenvalues at moderate coupling values. Calculation (a) fails to converge when the coupling becomes too strong because the diagonal preconditioning (diagonal of the uncoupled hamiltonian) becomes too crude. 
The quality is almost the same among all the eigenvalues at convergence. 
A total of 12, 13 or 14 correct digits are obtained for most eigenvalues. 
In a typical vibrational spectrum calculation with normal modes of the order of $1000$ cm$^{-1}$, the $12^{\text{th}}$ converged digit would roughly correspond to a $10^{-8}$ cm$^{-1}$ accuracy. 
\begin{figure}[htp]
\includegraphics[width=\linewidth]{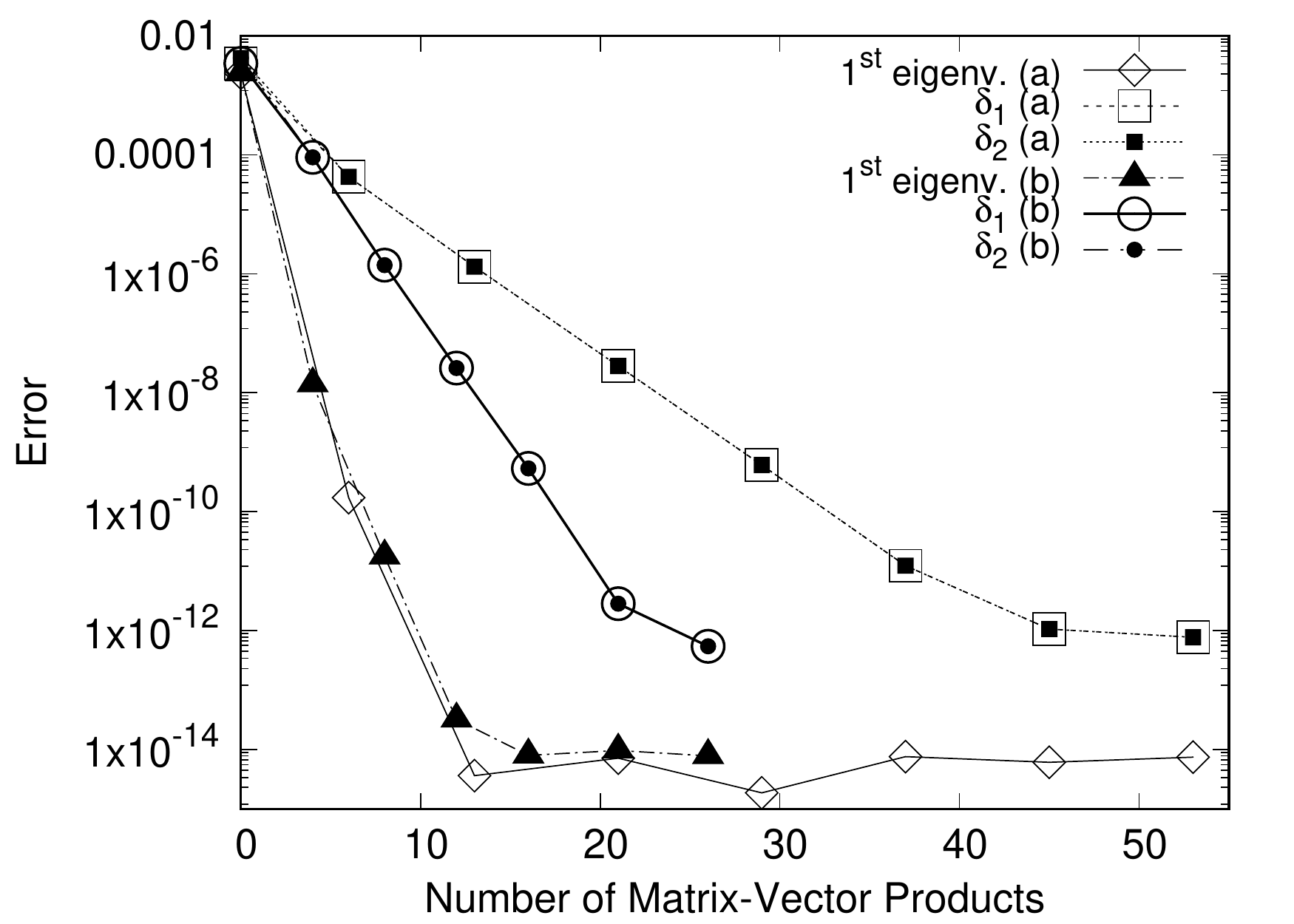}
 \caption{Convergence curves for calculations (a) using a diagonal preconditioner (squares) and (b) using the non-diagonal preconditioner (circles).  
 Results correspond to the moderate coupling case ($\veps = 0.080$). 
 The residuals $\delta_1$ and $\delta_2$ are calculated following Eqs.~\eqref{eqdelta1} and \eqref{eqdelta2} and are shown as a function of the number of matrix-vector applications per calculated vector. The convergence of the first eigenvalue is also shown (losanges and triangles for calculation (a) and (b), respectively). }
\label{fig_calc_a_b}
\end{figure}
The detailed numerical convergence of calculations (a) and (b) at moderate couplings ($\veps = 0.080$) are shown in Fig.~\ref{fig_calc_a_b}. 
As expected the convergence criteria $\delta_1$ and $\delta_2$ are equivalent and points are superimposed. 
Both calculations converge to the target accuracy within a few tens matrix-vector products. 
Calculation (a) reaches a global accuracy of $\delta_{1/2} =  7.7 \times 10^{-13}$ after 53 matrix-vector products and calculation (b) gives $\delta_{1/2} =  5.5 \times 10^{-13} $ after 26 matrix-vector products. 
The first eigenvalues converge faster than the higher-lying eigenvalues. 
Using the non-diagonal preconditioner (see appendix \ref{appprecond}) is more efficient, as expected.

\begin{table*}[ht]
\caption{Numerical parameters for the $4D$-coupled oscillator calculations and
their effect of the convergence of the adaptative wave operator algorithm. 
MVP means matrix-vector product per calculated eigenvector. }
\begin{tabular}{lccccc}
\hline
\hline
Calculation          			 	& (a)	& (b)	& (c)	& (d) 	& (e) \\
\hline 
Preconditioner						& diagonal 	& block-diag. & block-diag. & block-diag. & block-diag. \\
Intermediate subspace dimension $R$ & $-$ 	& 40	& 40 	& 100 	& 100 \\
Internal stopping criterion $\delta_{\text{stop}^{\text{int}}}$  & $10^{-1}$ & $10^{-1}$ & $10^{-2}$ & $10^{-1}$ & $10^{-2}$ \\
\hline 											
Number of MVP to reach convergence, with	&		&		&		&		&		\\
(i) weak coupling, $\veps = 0. 02 $	&	16	&	13	&	13	&	13	&	12	\\
(ii) moderate coupling, $\veps = 0.08$	&	53	&	26	&	24	&	20	&	21	\\
(iii) strong coupling, $\veps = 0.15 $	&	fails	&	45	&	45	&	34	&	39	\\
\hline 											
Convergence radius, $\veps_{max} =$	&	0.112	&	0.160	&	0.160	&	0.160	&	0.160	\\
\hline 											
\hline											
\end{tabular}
\label{tab_internal_par_effect}
\end{table*}

The first three rows of Table \ref{tab_internal_par_effect} show the internal numerical parameters used in five different calculations labelled (a) to (e). 
We discuss results with diagonal (a) or block-diagonal preconditioner (b-e), with different sizes for the intermediate subspace used in the block-preconditioner case ($R=40$ or $R=100$) and different values of the internal stopping criterion for the nested iterations $\delta_{\text{stop}^{\text{int}}}$. 
The remaining four rows in Table \ref{tab_internal_par_effect} show the effect of the internal numerical parameters on the convergence. We give the number of matrix-vector products (per calculated eigenvector) required to reach the convergence criterion $\delta_{1/2} = 10 ^{-12}$ at weak, moderate and strong coupling values. 
The method is not very sensitive to the stopping criterion of the nested iteration and so no particular effort is necessary to adjust this parameter. Actually comparison of columns (b) and (c) shows that the advantage is small to reduce $\delta_{\text{stop}^{\text{int}}}$. If smaller, the algorithm needs more nested iterations to find $Z^{(\ell)}$ but less iterations to refine the increment $\Delta Y^{(p)}$: at the end of the day the total number of matrix-vector applications is similar. 
It is clearly better to use the non-diagonal approximation for the preconditioning matrix (compare between (a) and (b), again). The convergence radius and speed are improved. 
Comparing (b) with (d) or (c) with (e), we see that increasing the size of the intermediate space $R$ gives an advantage especially at strong couplings but using a large intermediate space could rapidly become costly. Hopefully here the results with $R=2M$ are already good and not very different from results using $R=5M$, with only a few iterations more to converge.

\subsection{Comparison with standard methods}

In this subsection we compare results given by the adaptative wave operator method (AWO, section \ref{adaptativeAS}) with those given by the standard wave operator (SWO, section \ref{fixedAS}) method using a fixed active space or by a standard restarted block-Davidson method (B-DAV) \cite{davidson1975,liu1978,saad2011}. 
The main idea of the block-Davidson algorithm is to use projections of the hamiltonian matrix over a series of subspaces of increasing dimension. 
The basis set is progressively augmented by adding blocks of orthogonalized residuals of the eigenvectors associated with the lowest energies. 
The block-Davidson algorithm we use here is very similar to the one described in reference [\onlinecite{ribeiro2005}]. 
The initial block size (for B-DAV) and the active subspace (for AWO and SWO) have dimension $M=20$. 
We restart the B-DAV algorithm every 5 iterations to remain consistent, in terms of memory cost, with the AWO method where an intermediate subspace of dimension $R=100=5M$ has been selected.
\begin{table*}[ht]
\caption{Comparison of the adaptative wave operator algorithm (AWO) with the standard wave operator method (SWO) and with a restarted block-Davidson method (B-DAV).
MVP means matrix-vector product per calculated eigenvector. }
\begin{tabular}{lcccc}
\hline
\hline
Calculation          			 	& (e)	& (f)	& (g)	& (h) 	\\
\hline 
Method								& AWO	& SWO	& SWO	& B-DAV	\\
	Details	in						& Section \ref{adaptativeAS}
											& Section \ref{fixedAS} & Section \ref{fixedAS}
															& [\onlinecite{davidson1975,liu1978,ribeiro2005,saad2011}]  \\
Reuse of previous results			& yes	& no	& yes	& no 	 \\

Preconditioner						& bloc-diag. 	& diag. & diag. & diag. \\
Number of calculated eigenvectors 	& 20	& 20	& 20 	& 20 	\\
Dimension of the active space  $M$		& 20	& 20	& 20 	& - \\
Dimension of the intermediate space $R$ & 100 	& - 	& - 	& - \\
Restart every 						& 	-	&	-	&	-	& 	 5 \\
Maximum dimension of the projective space & - & - 	& - 	&  100 \\
\hline 
Number of MVP to reach $\delta_1 = 10^{-10}$ convergence	&		&		&		&			\\
(i) weak coupling, $\veps = 0. 02 $		&	12	&	10	&	9	&	26	\\
(ii) moderate coupling, $\veps = 0.08$	&	15	&	39	&	31	&	38	\\
(iii) strong coupling, $\veps = 0.15 $	&	21	&	fails	&	fails	&	56	\\
\hline
Cumulated number of MVP from $\veps = 0$ to...
										&		&		&		&		\\
(i) weak coupling, $\veps = 0. 02 $		&	80	&	79	&	73	&	154	\\
(ii) moderate coupling, $\veps = 0.08$	&	495	&	730	&	610	&	1084\\
(iii) strong coupling, $\veps = 0.15 $	&	1123&	-	&	-	&	2590\\
\hline
Convergence radius, $\veps_{max} =$		&	0.160	&	0.090	&	0.086	&	$>0.2$		\\
\hline 											
\hline											
\end{tabular}
\label{tab_comp_standard}
\end{table*}

\begin{figure}[htp]
\includegraphics[width=\linewidth]{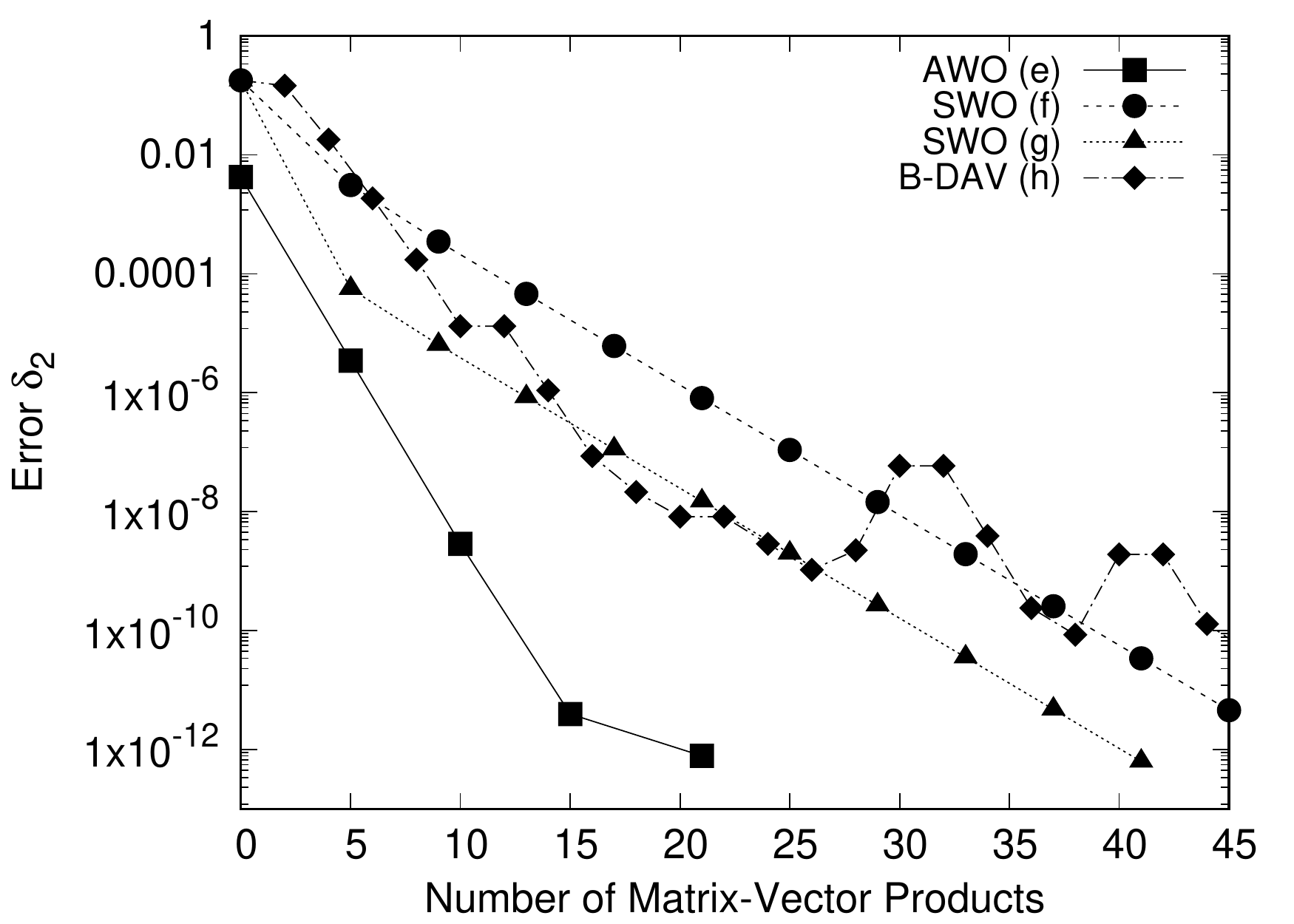}
 \caption{Convergence curves for calculations (e) to (h) described in text and in table \ref{tab_comp_standard}
 in the moderate coupling case ($\veps = 0.080$). 
The residual $\delta_2$ is calculated following Eq.~\eqref{eqdelta2} and is shown as a function of the number of matrix-vector applications per calculated vector. 
}
\label{fig_calc_efgh}
\end{figure}

Table \ref{tab_comp_standard} shows a comparison between four calculations.
Calculation (e) has been performed using the AWO algorithm 
with the same numerical parameters as in subsection \ref{sub_internal}.
Two calculations using the SWO 
are shown for comparison, using (g) or not using (f) previous results at each coupling iteration. 
Calculation (h) has been done with the B-DAV method. 
Together with different numerical parameters, the table gives the number of matrix-vector products required to reach convergence with $\delta_2 < \delta_{\text{stop}} = 10^{-10}$ at weak, intermediate and strong coupling values. The cumulated number of matrix-vector products is also given, i.e. the total number to reach convergence for all the coupling parameter values lower than the current one. 
In addition Fig. \ref{fig_calc_efgh} shows the detailed numerical convergence of the moderate coupling case ($\veps=	0.08$) for calculations (e) to (h). 
%
%

With weak coupling values, SWO and AWO are equivalently good and both perform better than B-DAV. 
When the coupling increases, the number of MVP for the AWO slightly increases but far less than for SWO, whose convergence becomes significantly slower. 
This is related to the fact that the Fubini-Study distance between the eigenspace and the adaptative active space always remains small using adaptative active suspaces, whereas this distance increases a lot in the case of SWO. 
As expected the SWO fails to converge when the eigenspace goes too far from the fixed active space (strong coupling case). 
The AWO do have a finite radius of convergence, even if it is far larger than the one of SWO calculations. 
AWO compares very well to B-DAV calculations.
Comparing (e) to (h), we see that AWO is almost two times faster than the B-DAV calculation. 
Moreover it should be noted that the B-DAV optimal accuracy does not exceed $\delta_2 \simeq 10^{-11}$, whereas accuracies better than $\delta_2 = 10^{-13}$ are obtained with AWO. 
The B-DAV could be rendered more efficient by restarting less frequently, but the memory advantage would be in favor of the AWO because the memory cost remains constant all along the calculation in wave operator calculations. 
We can conclude that the adaptative wave operator method performs well and is competitive with respect to a standard Block-Davidson algorithm in the present context of parameter-dependent hamiltonians.


\subsection{Stability of the algorithm with respect to dimensionality}

In this subsection we relax two assumptions made about the model hamiltonian of Eq.~\eqref{coupledhamiltonian}.
First we would like to check that the convergence of the method remains reasonable when the hamiltonian becomes larger so we set the dimensionality to $D=6$.
We have also relaxed the simple assumption of a single coupling constant (see Eq.~\eqref{eqsinglecoupling}), now defining the coupling constants between coordinates as 
\be 
\alpha_{ij} = \frac{1}{\vert i - j \vert}.
\label{eq:alphaijvar}
\ee 
Table \ref{tab_6D_parametres_coupled_osc} gives the numerical parameters used in the calculations. 
The direct product basis set has now dimension $N = 117\;649$. 
The active space dimension is still fixed at $M = 20$ and we use the non-diagonal preconditioner described in appendix \ref{appprecond} with an intermediate space of dimension $R=100$. 
\begin{table}[ht]
\caption{Numerical parameters for the $6D-$coupled oscillator calculations.}
\begin{tabular}{lr}
\hline
\hline
Number of $1D$ basis functions  	& $m_j=7$  \\
Total Hilbert space dimension 		& $N=117\,649$	 \\ 
Active subspace dimension 			& $M=20$ \\
Frequencies (arb. units)  &  $\omega_1 = \sqrt{2}$,  $\omega_2 = \sqrt{3}$ \\ 
& $\omega_3 = \sqrt{5}$,  $\omega_4 = \sqrt{7}$ \\ 
& $\omega_5 = \sqrt{11}$,  $\omega_6 = \sqrt{13}$ \\ 
Coupling parameter step  & $\Delta \veps = 0.002$  \\ 
Number of parameter steps  & $\mathscr{N} = 80$  \\  
\hline
\hline
\end{tabular}
\label{tab_6D_parametres_coupled_osc}
\end{table}
We ask for a global convergence criterion  $\delta_{1/2} < \delta_{\text{stop}} = 10^{-12}$. 
Results for the eigenvalues are given in Table \ref{tab_eigenvalues_coupledOH_6D} for two selected values of the coupling parameter, $\veps = 0.08$ and $\veps = 0.15$.
\begin{table}[ht]
\caption{Eigenvalues of the 6D-coupled oscillator model of Eq. \eqref{coupledhamiltonian} with couplings defined in Eq. \eqref{eq:alphaijvar}.  
The first column shows the global eigenvalue label, the second column contains the theoretical eigenvalues (exactly known here), the third columns show the numerical results given by the adaptative wave operator algorithm using numerical parameters  defined in table \ref{tab_6D_parametres_coupled_osc}. For each eigenvalue the last correct digit is written in bold type.
We show the results at two different values of the coupling parameter (moderate or strong). }
\begin{tabular}{ccc}
\hline	
\hline	
Label & 	Theoretical eigenvalues	&	$\quad$ Numerical results $\quad$ \\
\hline	
 \multicolumn{3}{c}{\textit{Coupling parameter}		 $\veps=	0.08$}\\
\hline	
1	 & 	7.47295046119813	&	7.472950461198{\bf{1}}6		\\
2	&	8.88121880840695	&	8.8812188084{\bf{0}}706	\\
3	&	9.20496110582695	&	9.204961105826{\bf{9}}9	\\
4	&	9.70729343955592	&	9.7072934395{\bf{5}}600	\\
5	&	10.11952829835106	&	10.11952829835{\bf{1}}15	\\
6	&	10.28948715561577	&	10.28948715561{\bf{5}}86	\\
\vdots & \vdots & \vdots \\
17	&	12.02149780024459	&	 12.0214978002{\bf{4}}516	\\
18	&	 12.19438185016026	&	12.19438185016{\bf{0}}41	\\
19	&	12.34524009766459	&	12.3452400976{\bf{6}}707	\\
20	&	12.35387127670885	&	12.3538712767{\bf{0}}902	\\
\hline							
 \multicolumn{3}{c}{	\textit{Coupling parameter} $\veps=	0.15$	}		\\
\hline							
1	 & 	7.46762124558304	&	7.46762124558{\bf{3}}12	\\
2	&	8.86234325858292	&	8.8623432585{\bf{8}}310	\\
3	&	9.19883617871454	&	9.19883617871{\bf{4}}72	\\
4	&	9.69758594506403	&	9.69758594506{\bf{4}}26	\\
5	&	10.11604954984208	&	10.11604954984{\bf{2}}27	\\
6	&	10.25706527158281	&	10.2570652715{\bf{8}}560	\\
\vdots & \vdots & \vdots \\
17	&	11.98828020471430	&	11.98828020{\bf{4}}93477	\\
18	&	12.16662262256184	&	12.1666226225{\bf{6}}204	\\
19	&	12.32477312484592	&	12.3247731{\bf{2}}575489	\\
20	&	12.34601424932307	&	12.34601424932{\bf{3}}51	\\
\hline 							
\hline 							
\end{tabular}
\label{tab_eigenvalues_coupledOH_6D}
\end{table}
All eigenvalues are well converged and other partial diagonalizations calculated along the coupling parameter discretization are also well converged. 
Similarly to the 4D case, convergence is respectively reached within
20 or 35
 matrix-vector products at moderate ($\veps = 0.08$) or strong ($\veps = 0.15$) coupling, respectively. 
This stability with respect to the dimensionality of the problem shows the potentialities of the method for molecular dynamics applications.

\section{Application to the calculation of Floquet quasienergy states for H$_2^+$ in intense laser fields \label{numerical2}}

\subsection{Floquet hamiltonian for H$_2^+$}

As a more realistic illustrative example, 
we wish to calculate the landscape of photodissociation resonances of a rotationless H$_2^+$ molecule submitted to a strong, linearly polarized laser field.  
We use a one-dimensional description within the framework of Born-Oppenheimer approximation,
with only two electronic states labelled $\vert 1\rangle$ and $\vert 2 \rangle$, corresponding to 
the ground electronic state X$^2\Sigma_g^+$ and to the purely repulsive excited state A$^2\Sigma _u^+$. 
The time-dependent hamiltonian can be written as
\bea
H (t)
 &= &
 T_N + 
\left[\begin{array}{c c} V_1(R)&0 \\
0&V_2(R) \end{array}\right] 
 - \mu_{12}(R)\mathcal{E}(t)
 \left[
 \begin{array}{c c} 0&1 \\
1&0 \end{array}
\right] \nonumber \\
&&
- i \left[ 
\begin{array}{c c} V_{\text{opt}}(R)&0 \\
0&V_{\text{opt}}(R) \end{array} 
\right]
\label{eqhamiltonian}
\eea
where
$T_N$ is the kinetic energy of the nuclei,
$V_1(R)$ and $V_2(R)$ are the Born-Oppenheimer potential energy curves associated with the two electronic states under consideration,
$\mu_{12}(R)$ is the electronic transition dipole moment between these two states \cite{bunkin}
and
$\mathcal{E}(t)$ is the electric field amplitude. 
The additionnal term $-iV_{\text{opt}}$ is a purely imaginary complex absorbing potential (optical potential), introduced to ensure that the outgoing wave Siegert boundary conditions are correctly reproduced in the discretized continua \cite{siegert, jolicardaustin}. 
The electric field is assumed to be a continuous wave, 
 \begin{equation}
 \mathcal{E}(t) = E\cos(\omega t)
 \end{equation}
with intensity $I\propto E^2$ and wavelength $\lambda= 2 \pi c /\omega$, $c$ being the speed of light. 
The Floquet theorem applies to this periodic coupling\cite{shirley},
and the dressed quasienergy states $\vert \chi_v \rangle$ are the eigenvectors of the Floquet eigenvalue problem defined as
\begin{equation}
H_{\text{F}}
\left[ \begin{array}{c} \chi_{1,v} (R,t)  \\
\chi_{2,v} (R,t) \end{array}\right]
= E_v \left[ \begin{array}{c} \chi_{1,v} (R,t)  \\
\chi_{2,v} (R,t) \end{array}\right]
\label{eq:eigenproblem}
\end{equation}
where the Floquet hamiltonian is 
\be 
H_{\text{F}} = \left[ H(t) - i \hbar \frac{\partial }{\partial t} \right].
\label{eqfloquethamiltonian}
\ee
In Eq.~\eqref{eq:eigenproblem}, $\chi_{1,v}$ and $\chi_{2,v}$ are the two components of the Floquet states associated with electronic states $1$ and $2$. We use label $v$ for the quasienergy states since they can be adiabatically linked to field-free vibrational states $\vert v\rangle$ from which they are originating. 
The associated resonance eigenvalues are complex,
\begin{equation}
E_v = \Re\text{e}(E_v)+i \Im\text{m}(E_v),
\label{eq:resonanceenergy}
\end{equation}
where $\Re\text{e}(E_v)$ is the energy and $\Gamma_v=-2 \Im\text{m}(E_v)$ is the resonance width or decay rate, inversely proportional to the lifetime. 

The above Floquet hamiltonian (Eqs.~\eqref{eqhamiltonian} and \eqref{eqfloquethamiltonian}) is a perfect example of a parametric hamiltonian. 
$H_{\text{F}}$ depends on two parameters $(I,\lambda)$ which can be discretized to explore a given laser intensity and wavelength domain and draw the quasienergy landscape. 
Two additional difficulties arise in comparison to the coupled oscillator problem:
(i) the Floquet hamiltonian  $H_{\text{F}}$ is non hermitian due to the complex absorbing potential and
(ii) many crossings or avoided crossings are present in the eigenvalue landscape due to the strong field coupling. 
The well-known presence of exceptional crossings between several eigenvalues will particularly retain our attention \cite{lefebvre2012}. Exceptional points (EP) are particular values of the parameters $(I,\lambda)$ where the Floquet hamiltonian becomes non-diagonalizable due to the self-orthogonality and coalescence of two eigenvectors\cite{moiseyevNHQM}. We will also explore the possibility of localizing zero-width-resonances in the quasienergy spectrum, i.e. particular combinations of the parameters leading to infinitely long-lived resonances \cite{atabek2006,leclerc2017}.

\subsection{Exploration of a selected eigensubspace in the 700-800~nm wavelength domain: zero-width resonances and exceptional crossings} 

We focus on the subspace originating from field-free vibrational states $v=11$ to $v=14$. To stabilize the calculation, the adaptative wave operator algorithm is run with a slightly larger active subspace of dimension $M=6$, initially chosen as the field free vibrational states $v=11,12,\dots,16$. 
In numerical calculations, the radial dependency is described using a standard Fourier DVR basis set with 200 functions for each electronic state. For the time-dependency we use a standard Fourier basis set made of $N_{F}$ functions of the form $e^{in\omega t}$ with $n= -N_F/2, \dots, N_F/2-1 $. We take into account $N_F=4$ Floquet blocks to accurately describe all the multiphoton processes. 
The Floquet hamiltonian is not very large ($N=1600$) but it is not as sparse as was the coupled oscillator hamiltonian of section \ref{numerical1} %
and it is no longer hermitian. 
 We consider a two-dimensional grid of $\mathscr{N} = 100\times 100 = 10^4$ discrete values of the laser parameters $(I,\lambda)$ and we apply the adaptative wave operator algorithm along lines of increasing intensity at fixed wavelength.

To build an efficient non-diagonal preconditioner (see Appendix \ref{appprecond}), the states participating in the intermediate subspace are selected as follow.
After the calculation at a given intensity is converged, we inspect the $M$ eigenvectors and we identify for each one of them the leading components, {\textit{i.e.}} the states contributing the most to the eigenvectors (excluding those already present in the active space). We select $12$ states contributing the most to each of the $M$ individual eigenvectors and we include them sequentially in the intermediate subspace (always avoiding repetitions), and finally we add $12$ states on the basis of their global weight along all $M$ eigenvectors, taken as a whole. The intermediate space has thus a total dimension of $R = 90$. At the very first iteration of the intensity, the selection of the intermediate subspace cannot be based on such weighting criteria, so we select the $R$ states closest in energy from the active subspace.

\begin{figure*}[htp]
\includegraphics[width=0.45\linewidth]{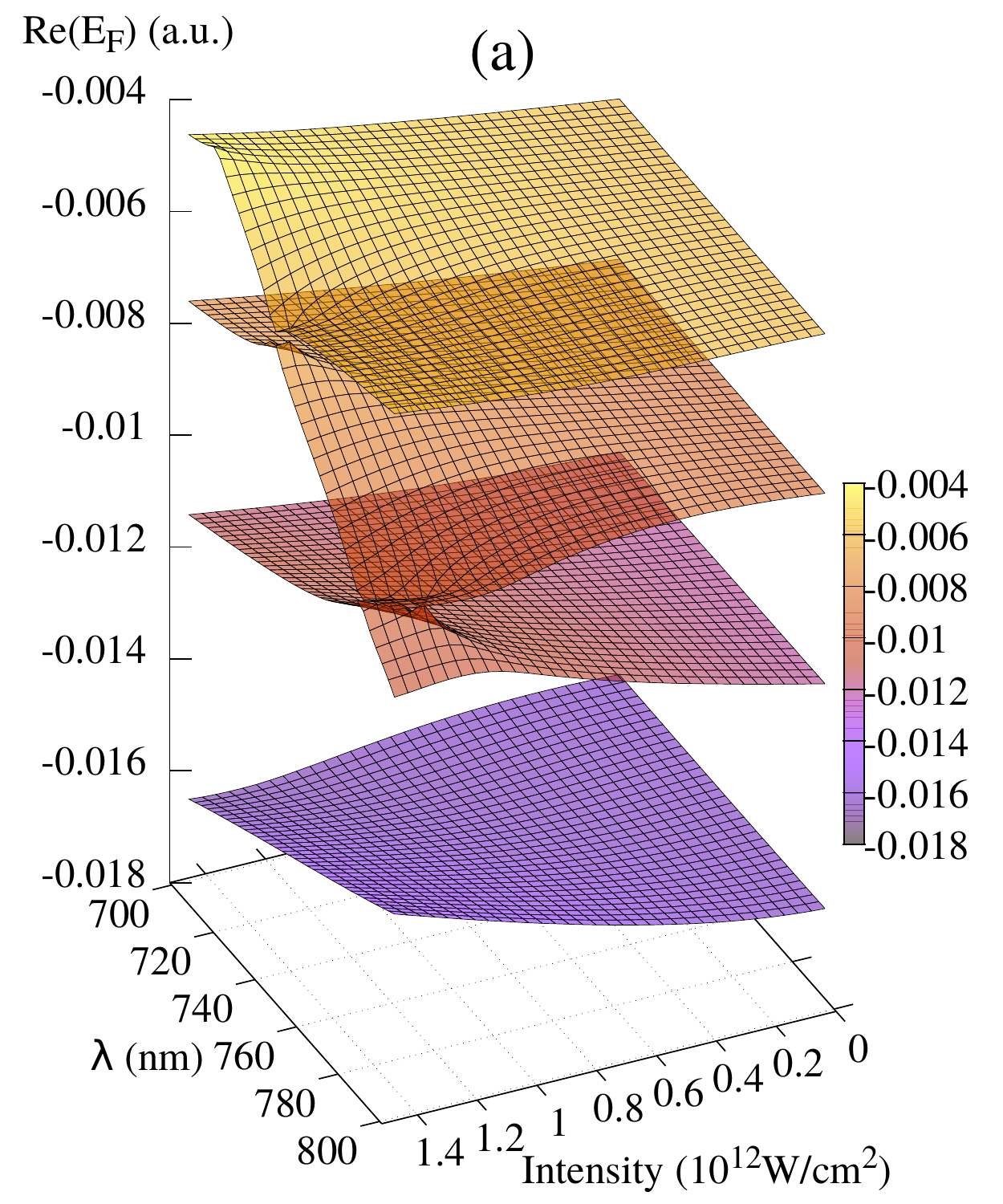}
\includegraphics[width=0.45\linewidth]{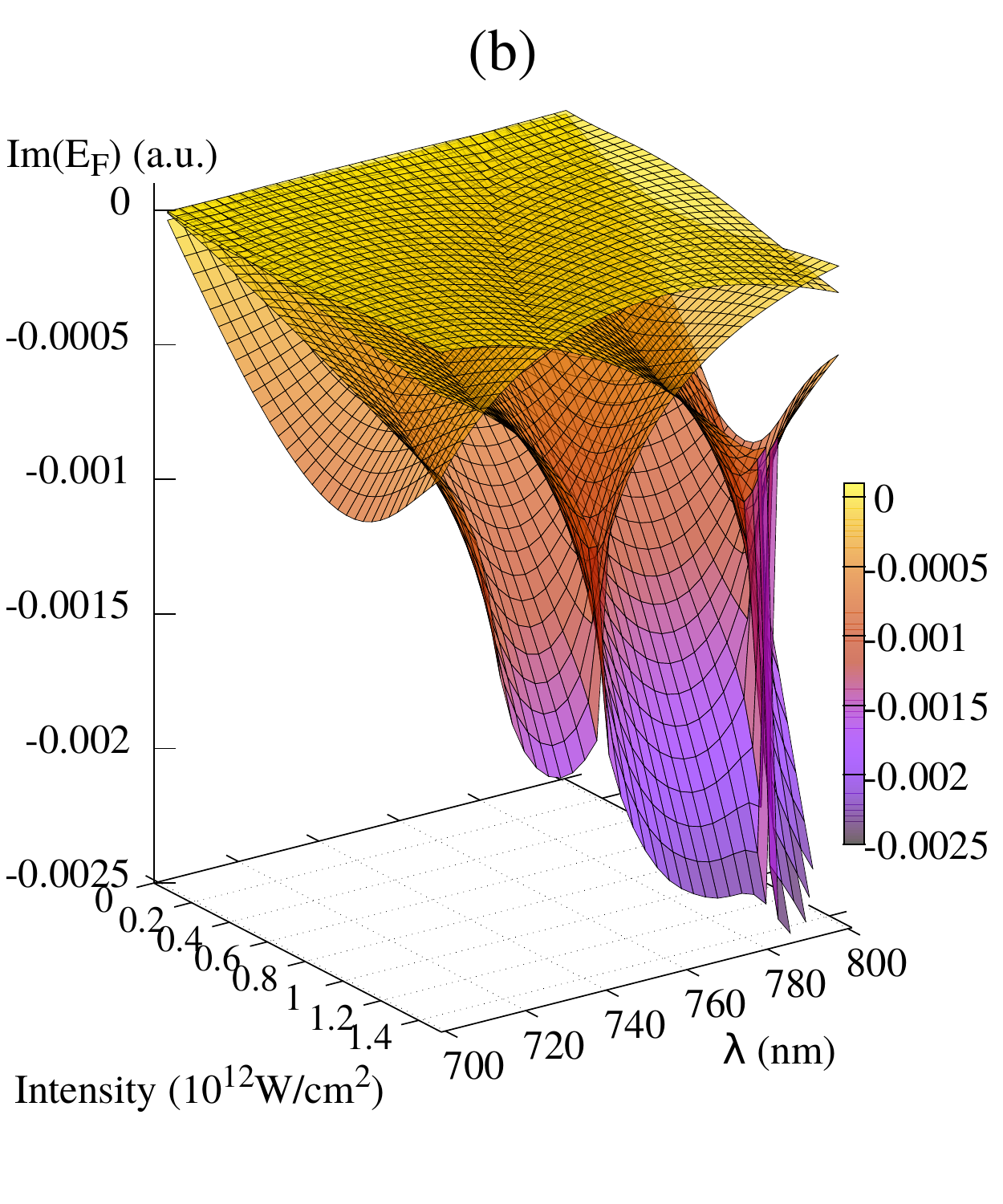}
 \caption{Floquet eigenvalue surfaces for photodissociation resonances originating from field-free vibrational states $v=11$ to $v=14$, calculated using the adaptative wave operator algorithm over a wavelength domain $\lambda \in [700 \text{nm},800 \text{nm}]$ and for an intensity between $0$ and $I_{\text{max}} = 1.5 \times 10^{12}$~W.cm$^{-2}$. Real parts are shown in panel (a) and imaginary parts in panel (b). 
}
\label{fig_floquet_surface}
\end{figure*}

We ask a $\delta_{\text{stop}} = 10^{-5}$ convergence and the internal stopping criterion is set at $\delta_{\text{stop}^{\text{int}}} = 0.1$. 
The adaptative algorithm needs no more than 12 matrix-vector products to reach convergence, except in the vicinity of crossings where the convergence is sometimes a bit slower. 
The main results are shown in Fig. \ref{fig_floquet_surface}. Starting at real values associated with the field-free vibrational states $v=11$ to $v=13$, the dressed eigenvalues appears as Riemann sheets with several crossings of their real parts as well as their imaginary parts.

\begin{figure*}[htp]
\includegraphics[width=0.42\linewidth]{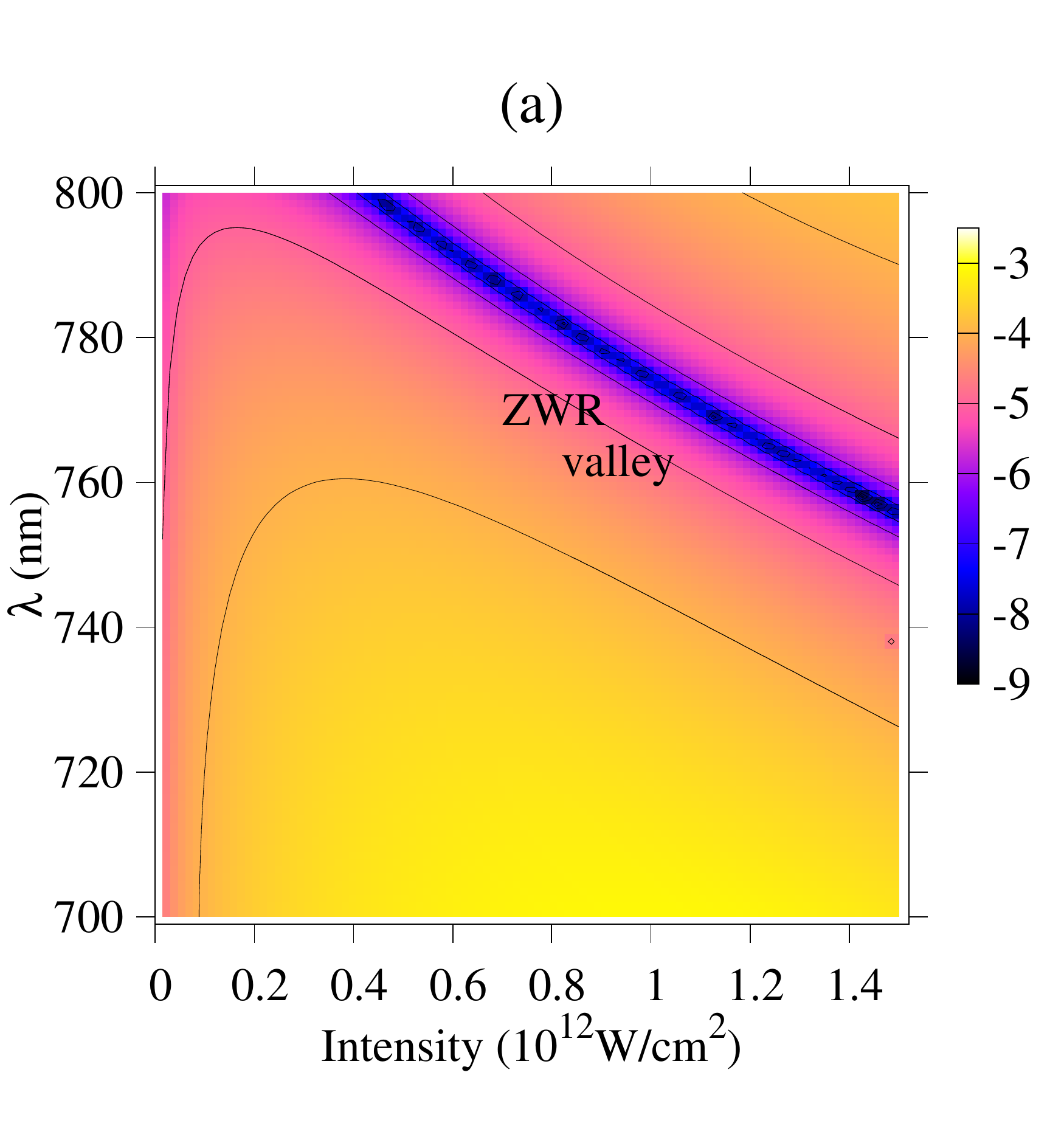}
\includegraphics[width=0.42\linewidth]{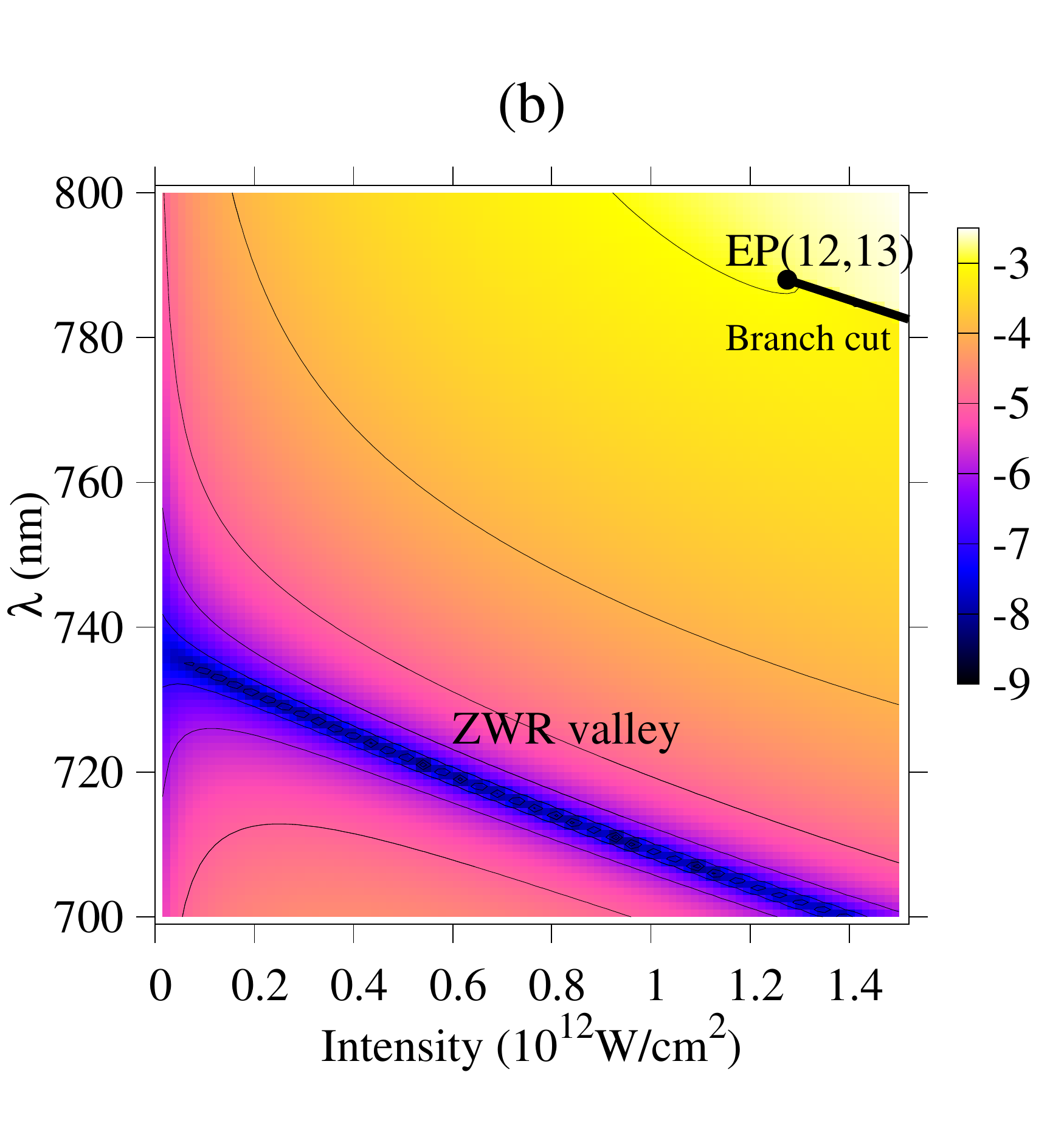}
\includegraphics[width=0.42\linewidth]{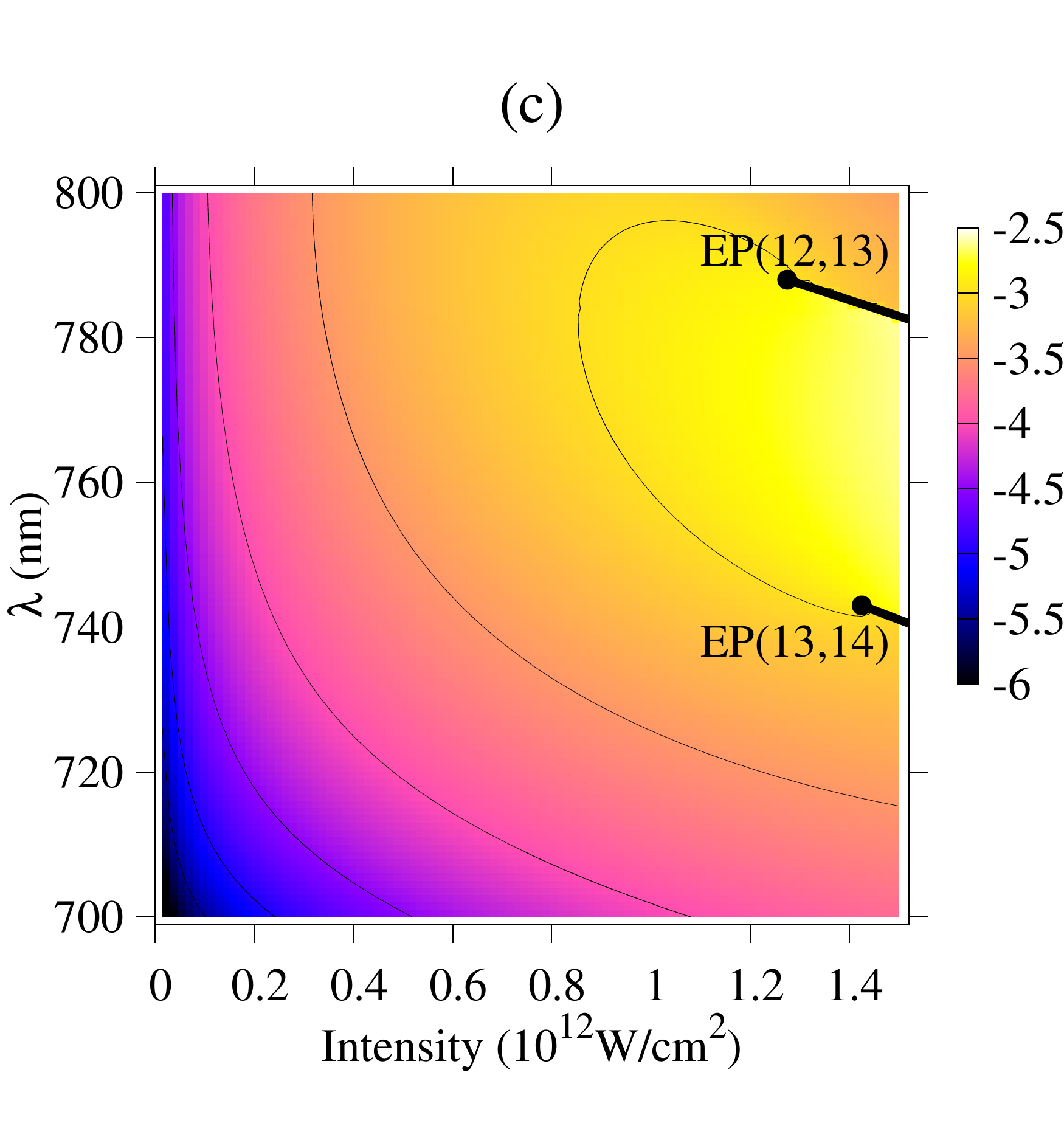}
\includegraphics[width=0.42\linewidth]{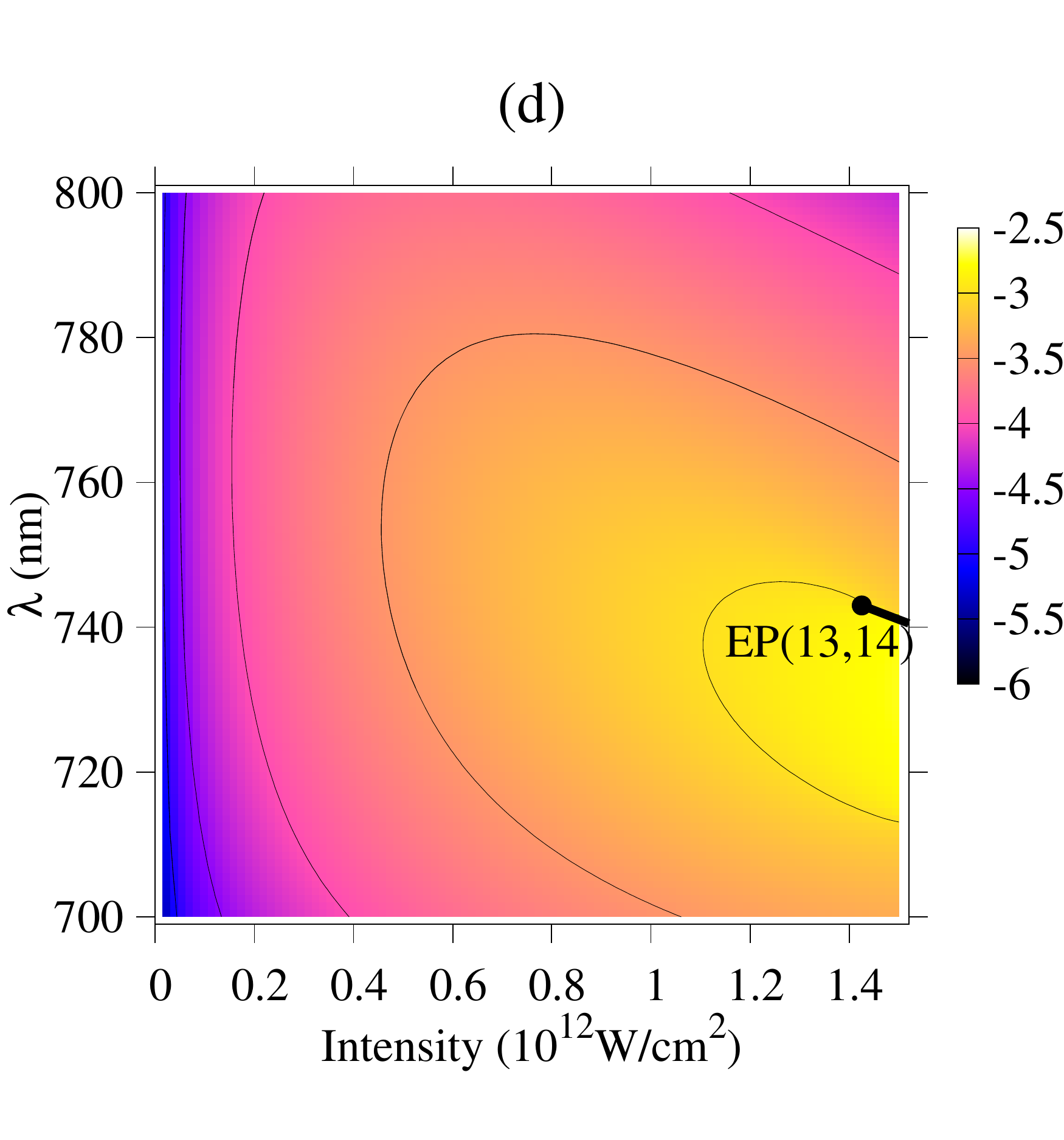}
 \caption{Map of the resonance widths in logarithmic scale. The color scale represents $\log_{10}(|\Im\text{m}{(E_v)}|) =\log_{10}(\Gamma /2)$ in the intensity-wavelength parameter plane. Panels (a) to (c) correspond to resonances originating from field-free vibrational state $v=11$ to $v=14$, respectively. The blue lines in panels (a) and (b) are zero-width resonance valleys. Branch cuts are indicated as black lines and two exceptional points are visible: EP(12,13) in panels (b) and (c), and EP(13,14) in panels (c) and (d).}
\label{fig_width}
\end{figure*}

To highlight these crossings and clarify the results for the imaginary part (\textit{i.e.} the resonance width), Fig. \ref{fig_width} shows the four resonance widths in logarithmic scale. Zero-width resonance regions are clearly identified on the first two panels (a) and (b) associated with field-free states $v=11$ and $v=12$. Their quasi-linear behavior in the $(I, \lambda)$ parameter plane is consistent with linear approximations suggested in previous works\cite{atabek2013,leclerc2017}. Exceptional points are also well identified on panels (b), (c) and (d), between resonances originating from $v=12,13$ and $v=13,14$. The results are consistent with the cluster of exceptional points expected for H$_2^+$ in this wavelength domain from the work of Lefebvre \textit{et al}\cite{lefebvre2012}.

\section{Conclusion}

We have developed an adaptative wave operator algorithm designed to efficiently extract a few eigenvalues and their associated eigenvectors from large parameter-dependent matrices. 
Working equations have been derived to make the active subspace  follow the eigenspace as close as possible when coupling terms in the matrix are increasingly modified. 
The equations are a little more complicated than in the static, standard wave operator algorithm, but as in every iterative method the simplicity comes from the fact that only matrix-vector products and projections are needed. 
%

%

Convergence difficulties at strong coupling values are the usual weakness of wave operator algorithms. Such difficulties are not completely removed by the use of adaptative active subspaces alone. This is also dealt with by using non-diagonal preconditioning matrices defined in a subspace of intermediate dimension, which can be either selected at the begining of the calculation, or can be progressively updated when the coupling is increased.

A full numerical investigation has been carried out using a model hamiltonian describing an ensemble of coupled oscillators. The adaptative algorithm is far more efficient than the standard wave operator algorithm and it compares very well to the standard restarted Davidson algorithm. 
The method has also been applied to H$_2^+$ photodissociation, calculating quasienergy states of the non-hermitian Floquet hamiltonian in a selected intensity-wavelength domain and drawing maps of resonance widths. This feature may be very interesting to explore the parameter space in quantum control problems. 

The idea of an adaptative subspace and the associated working equations could also be applied to any eigenproblem where an approximate solution is already known, even if their are no physically varying parameters. 
In this way, the adaptative wave operator method could be used to build iteratively a good active space to diagonalize a given Hamiltonian as those arising, for example, in large scale vibrational spectra calculation\cite{carrington2017}, in the same spirit as what has been proposed in refs [\onlinecite{scribano2008,garnier2016,lesko2019}], but at constant memory cost.

\appendix

\section{Projective representation of the Bloch equation \label{AppCalc}}

In this short appendix we simplify Eq.~\eqref{BWOn}-\eqref{eqHeffn} to avoid $N\times N$ matrices. 
Equation \eqref{BWOn} can be expanded as
\be 
( P_n +Q_n X_n P_n) H_n (P_n + Q_n X_n P_n) = H_n (P_n + Q_n X_n P_n).
\ee 
Projecting on the left on $Q_n$ leads to
\be 
Q_n X_n P_n H_{\text{eff},n} P_n = Q_n H_n P_n + Q_n H_n Q_n X_n P_n
\ee 
or equivalently
\bea
&Q_n &X_n \Vc_{n-1} \Vc_{n-1}^{\dagger} H_{\text{eff},n} \Vc_{n-1} \Vc_{n-1}^{\dagger}  \nonumber \\
&=&
Q_n H_n \Vc_{n-1} \Vc_{n-1}^{\dagger} + Q_n H_n Q_n X_n \Vc_{n-1} \Vc_{n-1}^{\dagger}. 
\eea
Closing with $\Vc_{n-1}$ on the right and using the normalization condition \eqref{normalizationV}, we get the following more compact formula:
\be 
Y_n \mathcal{M}_{\text{eff},n} = Q_n H_n \Vc_{n-1} + Q_n H_n Q_n \; Y_n,
\ee
where the unknowns are
\be 
Y_n  \equiv Q_n X_n \Vc_{n-1},
\ee 
and 
\be 
\mathcal{M}_{\text{eff},n} \equiv 
\Vc_{n-1}^{\dagger} H_n \Omega_n \Vc_{n-1}. 
\ee 

\section{Fixed point iteration using diagonal preconditioning matrices \label{AppZ} }

In this appendix we give some details about the internal iterative solution of Eq.~\eqref{eqZ} in the case of diagonal preconditioning matrices. 
As a preconditioning step we add on each side of Eq.~\eqref{eqZ} the quantity $(ZD' - H'Z)$ with $D'$ and $H'$ two arbitrary diagonal matrices of respective sizes $M\times M$ and $N\times N$. 
This leads to 
\be 
ZD' - H'Z = \mathcal{F}(Y_n^{(p-1)}) + (Q_n H_n - H') Z - Z(D_n-D'),
\label{eqZHprime}
\ee 
or equivalently, written in elementwise notation:
\be 
Z_{ij} = 
\frac{
\left[ \mathcal{F}(Y_n^{(p-1)}) + (Q_n H_n - H') Z - Z(D_n-D')  \right]_{ij}
}{
D'_{jj} - H'_{ii}
}
\label{eqZij}
\ee 
for $i=1,\dots, N$ and $j=1,\dots,M$.
This is a self-consistent equation of the form  $ Z = \mathcal{G} (Z) $ for the unknown $Z=\Delta Y \Tc_n$ .
From the fixed-point theorem, it can be used to build a nested iteration for $Z$. 
If it converge, the sequence defined as $Z^{(\ell)} = \mathcal{G} ( Z^{(\ell-1)} )$ converge to the solution of Eq.~\eqref{eqZij}. 
Starting with a zero $N \times M$ matrix for $Z^{(0)}$, $Z$ can be found by building the internal sequence
\begin{widetext}
\be 
Z_{ij}^{(\ell)} = 
\frac{
\left[ \mathcal{F}(Y_n^{(p-1)}) + (Q_n H_n - H') Z^{(\ell-1)} - Z^{(\ell-1)}(D_n-D')  \right]_{ij}
}{
D'_{jj} - H'_{ii}
}, \quad \ell = 1,\dots,\mathscr{L}
\label{eqZijiterationapp}
\ee 
\end{widetext}
for $i=1,\dots, N$ and $j=1,\dots,M$.

\section{Error propagation in the adaptative wave operator internal iterative equation \label{appconvergence}}

To ensure the convergence of the iterative procedure defined in Eq.~\eqref{eqZijiteration} for the increment $Z$, the $D'$ and $H'$ preconditioning matrices should be chosen reasonably close to $D_n$ and $Q_n H_n$, respectively. To justify this recommendation, let us consider Eq.~\eqref{eqZijiteration} written in line as
\bea
Z^{(\ell)}D' - H'Z^{(\ell)} &=& \mathcal{F}(Y_n^{(p-1)})+ (Q_n H_n - H') Z^{(\ell-1)} \nonumber  \\
 &-& Z^{(\ell-1)}(D_n-D').
\label{eqZl}
\eea
The exact solution of Eq.~\ref{eqZHprime} satisfies 
\bea
Z^{(\text{ex})}D' - H'Z^{\text{(ex)}} &=& \mathcal{F}(Y_n^{(p-1)}) + (Q_n H_n - H') Z^{\text{(ex)}} \nonumber \\
 &-& Z^{\text{(ex)}}(D_n-D'). 
\label{eqZex}
\eea
Suppose that the matrices of absolute errors on $Z$ at steps $(\ell)$ and  $(\ell-1)$ are respectively written as
\be
\delta_Z^{(\ell)} = Z^{(\ell)} - Z^{(exact)}
\ee 
and
\be 
\delta_Z^{(\ell-1)} = Z^{(\ell-1)} - Z^{(exact)}.
\ee 
Then by substracting Eq.~\eqref{eqZex} from Eq.~\eqref{eqZl}, we obtain
\be 
\delta_Z^{(\ell)} D' - H'\delta_Z^{(\ell)} = (Q_n H_n - H') \delta_Z^{(\ell-1)} - \delta_Z^{(\ell-1)}(D_n-D').
\label{eqerror}
\ee 
The error at step $(\ell-1)$ is propagated at step $(\ell)$ following the elementwise relationship
\be 
\left[{\delta_Z^{(\ell)}}\right]_{ij} =
\frac{
\left[ 
(Q_n H_n -H') \delta_Z^{(\ell-1)} - \delta_Z^{(\ell-1)} (D_n-D')
\right]_{ij}
}
{
D'_{jj} - H'_{ii}
}
\label{eqerrorprop}
\ee 
for  $i=1,\dots, N$ and $j=1,\dots,M$. 
If the numerator of \eqref{eqerrorprop} becomes too large during the iterations, or if some elements of the denominator become too small, the error can accidentally grow and convergence can be lost. The convergence radius is thus mainly controlled by the difference between the exact $Q_n H_n$ operator and the preconditioning matrix $H'$, and in a lesser extent by the difference between $D_n$ and $D'$. 

\section{Non-diagonal preconditioner for the adaptative wave operator algorithm \label{appprecond}}

In this appendix we show how the working equations of the nested internal iteration described in subsection \ref{internal} and appendix \ref{AppZ} are modified when using a non-diagonal preconditioning matrix $H'$.
We mainly refer the reader to Eqs.~\eqref{eqZHprime}, \eqref{eqZij} and \eqref{eqZijiteration} before reading this appendix. 
Here we assume that a non-diagonal $H'_{\text{n.d.}}$ preconditioning matrix is introduced in Eq.~\eqref{eqZHprime}. Indeed the convergence of the nested iteration \eqref{eqZijiteration} strongly depends on the difference between $Q_n H_n Q_n$ and $H'$. The smaller the difference, the better the convergence. 
We also assume that $H'_{\text{n.d.}}$ remains easily diagonalizable. 
A clever choice is to define $H'_{\text{n.d.}}$ as an intermediate block matrix of size $R \times R$ with $M < R \ll N$ containing the matrix representation of $ Q_n H_n Q_n$ within what we call the ``intermediate ($R-$dimensional) subspace". The selection of which basis vectors may form the intermediate subspace may be based on physical considerations (for example an energy cut-off of the harmonic oscillator basis set in a vibrational problem). 
The rest of the $H'_{\text{n.d.}}$ matrix is kept diagonal, with the remaining $(N-R)$ elements filled in by the corresponding diagonal elements of $H_n$. 
The non-diagonal preconditioner is thus:
\be 
H'_{\text{n.d.}} = 
\left(  
\begin{array}{c|c}
(Q_n H_n Q_n) _{R\times R} & 0 \\
 \hline 
 0 & \text{diag}(H_n) \\
\end{array} 
\right) 
\label{eqprecondnd}
\ee 
with
\be 
M < R \ll N. 
\ee
If $R$ is small, say a few times $M$, then $H'_{\text{n.d.}}$ can be easily block-diagonalized,
\be 
(Q_n H_n Q_n) _{R\times R} = S_R \Lambda_R S^{-1}_R,
\ee
with $\Lambda_R$ a $R\times R$ diagonal matrix. Then the preconditioner diagonalization is
\be 
H'_{\text{n.d.}} = S \Lambda S^{-1}
\ee 
with $\Lambda$ a $N\times N$ diagonal matrix given by 
\be 
\left(  
\begin{array}{c|c}
\Lambda_R & 0 \\
 \hline 
 0 & \text{diag}(H_n) \\
\end{array} 
\right)
\ee 
and $S$ the $N\times N$ associated eigenvector matrix:
\be 
S = 
\left(  
\begin{array}{c|c}
S _R & 0 \\
 \hline 
 0 & I_{N-R} \\
\end{array} 
\right).
\ee 
Eq.~\eqref{eqZHprime} becomes 
\bea
&ZD'& - S \Lambda S^{-1} Z  \nonumber \\
&=& \mathcal{F}(Y_n^{(p-1)}) 
  + (Q_n H_n - S \Lambda S ^{-1}) Z \nonumber \\
&& - Z(D_n-D').
\label{eqZApp}
\eea
Left multiplication of Eq.~\eqref{eqZApp} by $S^{-1}$ gives a new self-consistent equation for the $N \times M$ matrix $\eta$ defined by
\be 
\eta \equiv S^{-1} Z,
\ee 
which satisfies the following equation:
\begin{widetext}
\be
\eta D' - \Lambda \eta = S^{-1} \mathcal{F}(Y_n^{(p-1)})
+ S^{-1} Q_n H_n S \eta  - \Lambda \eta - \eta (D_n-D'). 
\label{eqeta}
\ee
Eq.~\eqref{eqeta} 
 can be iterated in a similar way to Eq.~\eqref{eqZijiteration}, 
\be 
\eta_{ij}^{(\ell)} = 
\frac{
\left[ S^{-1} \mathcal{F}(Y_n^{(p-1)}) + S^{-1} Q_n H_n S \eta^{(\ell-1)} - \Lambda \eta^{(\ell-1)} - \eta^{(\ell-1)}(D_n-D')  \right]_{ij}
}{
D'_{jj} - \Lambda_{ii}
}, \quad \ell = 1,\dots,\mathscr{L}
\label{eqetaijiteration}
\ee 
\end{widetext}
for $i=1,\dots, N$ and $j=1,\dots,M$.
To ensure that the projective property \eqref{eqprojYn} remains satisfied, we project onto the complementary subspace between two successive iterations, 
\be 
\eta^{(\ell)} \leftarrow Q_n \eta^{(\ell)}.
\ee 
After convergence, $Z$ is recovered by simply forming $Z^{(\mathscr{L})} = S \eta^{(\ell=\mathscr{L})}$ and the increment is calculated using Eq.~\eqref{calculDeltaY}.

\begin{acknowledgments}
The PMMS (P\^ole Messin de Mod\'elisation et de Simulation) is gratefully acknowledged for providing us with computer time. 
Some of the calculations have been executed on computers of the UTINAM Institute at the Universit\'e de Franche-Comt\'e, supported by the R\'egion de Bourgogne Franche-Comt\'e and Institut des Sciences de l'Univers (INSU).
This research was supported by CNRS GDR 3575 THEMS. 
\end{acknowledgments}


\end{document}